\documentclass{article}
\usepackage[utf8]{inputenc}
\usepackage{amsmath,amssymb,amsfonts,stmaryrd}
\usepackage{physics}
\usepackage{dsfont}
\usepackage{graphicx}
\usepackage{xcolor}
\usepackage{graphicx}
\usepackage{cancel}
\usepackage{hyperref}
\hypersetup{
    colorlinks=true,
    linkcolor=black,
    filecolor=red,      
    urlcolor=blue,
    citecolor=red
    }
\usepackage{bm} %allows for bold math type
\usepackage{subfigure} %allows for 2x2 figures
\usepackage{environ}
\usepackage{url}
\usepackage[margin=0.75in]{geometry}
\usepackage[mathscr]{euscript}
\usepackage[scr=boondox]{mathalpha}

\newcommand{\Z}{{\mathbb Z}}

\NewEnviron{eqs}{%
\begin{align}\begin{split}
    \BODY
\end{split}\end{align}
}

\title{
Generalized Hall Conductivities in Local Commuting Projector Models: Generalized Symmetries and Protected Surface Modes
}
\author{Po-Shen Hsin$^{1}$, Ryohei Kobayashi$^{2}$}
\date{\today}

\begin{document}

\maketitle

\begin{center}

${}^1$ Department of Mathematics, King’s College London, Strand, London WC2R 2LS, UK \\
 ${}^2$School of Natural Sciences, Institute for Advanced Study, Princeton, NJ 08540, USA
 \end{center}

\bigskip\bigskip
\begin{abstract}
Hall conductivities are important characterizations of phases of matter. 
It is known that nonzero Hall conductivities are difficult to realize in local commuting projector lattice models due to no-go theorems in (2+1)D.
In this work we construct local commuting projector models in (2+1)D and (3+1)D with nonzero generalized Hall conductivities 
for ordinary and higher-form continuous symmetries on tensor product Hilbert space of finite local dimension. The model is given by a standard $\mathbb{Z}_N$ toric code, but the symmetries do not admit expression in terms of onsite charge operators. The symmetry do not have local charges or currents on the lattice in the absence of boundaries, but there is still notion of Hall conductivities that coincide with the continuum field theories. We construct protected gapless boundaries of the lattice models using modified Villain formalism.
The generalized Hall conductivities are computed by surface currents as well as bulk flux insertion and many body Chern number.

\medskip
\noindent
\end{abstract}

\bigskip \bigskip \bigskip
 
% \date{\today}

\bigskip

% \eject

\tableofcontents

\unitlength = .8mm

\setcounter{tocdepth}{3}

\bigskip

\section{Introduction}
\label{sec:intro}

Hall conductivities underlie many experimental breakthroughs such as fractional quantum Hall (FQH) systems, e.g.~\cite{Park_2023,Lu2024} for recent experiments. The Hall conductivity $\sigma_{xy}$ relates electric charge $Q$ to the magnetic flux $\Phi$ through regions in 2d space,
\begin{equation}\label{eqn:Hallordinary}
    Q=\frac{\sigma_{xy}}{2\pi}\Phi~.
\end{equation}
The Hall conductivity indicates that the boundary has protected gapless modes that carry current for the conductivity. 
While Hall conductivity is important, it is known that Hall conductivity must vanish in local commuting projector models with onsite $U(1)$ symmetry and finite dimensional local Hilbert space \cite{Kapustin_2019,Zhang2022hall}, with similar result for thermal Hall conductivity \cite{Kitaev_2006,Kapustin_2020}.
Thus while local commuting projector models are useful in describing many phenomena such as topological orders, e.g.~\cite{Levin_2005,walker201131tqftstopologicalinsulators,Chen_2013,PhysRevA.83.042330}, it remains challenging to describe systems with nontrivial Hall conductivities.

There are various attempts to evade the no-go theorems~\cite{DeMarco:2021erp,sengoku2025quasilocalfrustrationfreefreefermions,Han_2022}. One way for evasion is to use infinite dimensional Hilbert space, such as using rotor degrees of freedom \cite{DeMarco:2021erp}, or treating the bulk degrees of freedom as infinitely many ``flavor'' degrees of freedom on each site \cite{Hsin:2022iug}.\footnote{
For example, if we regard the Walker-Wang models \cite{walker201131tqftstopologicalinsulators} for chiral anyon theories as (2+1)D theory with infinite number dimensional local Hilbert space labeling the bulk coordinate direction \cite{Hsin:2022iug}, then the Walker-Wang models provide local commuting models with nonzero Hall conductance and thermal Hall conductance \cite{Kobayashi2024U1WW, Wang2022exactly}. 
} Another way for evasion is to admit exponentially decaying tails in the local Hamiltonians, and then a frustration free Hamiltonian can host nontrivial Chern number for $U(1)$ symmetry \cite{sengoku2025quasilocalfrustrationfreefreefermions}.
However, it was unclear whether the no-go theorems can be evaded with finite dimensional local Hilbert space with exact locality.

In this work we will construct local commuting projector lattice models with finite dimensional local Hilbert space and nontrivial generalized Hall conductivities. This shows that infinite dimensional local Hilbert space or decaying tails in local Hamiltonian is not essential for non-vanishing Hall conductivities in local commuting models.

\subsection{Generalized fractional quantum Hall states with local commuting projector model}

The models we will use are $\mathbb{Z}_N$ toric codes \cite{KitaevAnyon}, which have continuum limit given by $\mathbb{Z}_N$ gauge theories enriched with $U(1)_E\times U(1)_M$ symmetries in various dimensions:
\begin{equation}\label{eqn:continuum}
    \frac{N}{2\pi}adb+\frac{1}{2\pi}adA_E+\frac{1}{2\pi}bdA_M~,
\end{equation}
where $a,b$ are dynamical gauge fields, and $A_E,A_M$ are background gauge fields for the $U(1)$ symmetries. The theory is a generalization of the Abelian Chern-Simons theory description of fractional quantum Hall states, e.g. \cite{PhysRevB.42.8133}, where the gauge fields can be higher-form instead of one-form, and the symmetries can be higher-form symmetries \cite{Gaiotto:2014kfa}. 
The corresponding conserved currents are
\begin{equation}
    J_E=\frac{1}{2\pi}\star da,\quad J_M=\frac{1}{2\pi}\star db~,
    \label{eq:u1currents}
\end{equation}
where $\star$ is the Hodge star operation. The symmetries are generated by the operators given by theta terms:
\begin{equation}
    U_\theta^e(\Sigma)= e^{i\frac{\theta}{2\pi}\int_\Sigma da},\quad  
    U_\theta^m(\Sigma)= e^{i\frac{\theta}{2\pi}\int_\Sigma db}~,
\end{equation}
where $\theta$ parameterize the $U(1)$ transformations. These operators are trivial when $\Sigma$ is a closed oriented submanifold, but they can form nontrivial junctions between different transformations $\theta$: for example, the junction of $U^e_{[\theta_1]},U^e_{[\theta_2]},U^e_{[\theta_1+\theta_2]}$ (we will use $[\cdot]$ to denote the restriction of angular parameter to $[0,2\pi)$) is the line operator
\begin{equation}\label{eqn:junctionsym}
    \alpha_{\theta_1,\theta_2}=e^{iq(\theta_1,\theta_2) \int a},\quad q(\theta_1,\theta_2)=\frac{[\theta_1]+[\theta_2]-[\theta_1+\theta_2]}{2\pi}\in\mathbb{Z}~.
\end{equation}
The nontrivial junction is the hallmark of symmetry fractionalization, where the line operator $\alpha_{\theta_1,\theta_2}$ at the junction describes the vison that detects fractional charge of line operators that braid with the junction \cite{Barkeshli:2014cna,Teo_2015,Tarantino_2016,Benini:2018reh}.

As in fractional quantum Hall states, the continuum field theory description (\ref{eqn:continuum}) for the models can be shown to have fractional generalized Hall conductivity given by the effective Chern-Simons like response action
\begin{equation}
    \frac{\xi}{2\pi} A_MdA_E,\quad \xi=-\frac{1}{N}~.
    \label{eq:generalhall}
\end{equation}

For $D=3$ spacetime dimension and one-form $\mathbb{Z}_N$ gauge theory, the symmetries are $U(1)_M^{(0)}\times U(1)_E^{(0)}$ ordinary symmetries, and the response is the fractional quantum Hall conductivity with $\sigma_{xy}=-\frac{1}{N}$. 
\begin{equation}
    Q=\frac{\sigma_{xy}}{2\pi}\Phi,\quad \sigma_{xy}=-\frac{1}{N}~.
\end{equation}

For $D=4$ spacetime dimension and one-form $\mathbb{Z}_N$ gauge theory, the symmetries are $U(1)_M^{(0)}\times U(1)_E^{(1)}$  ordinary and one-form symmetries, and the response is the generalized fractional quantum Hall conductivity with $\xi=-\frac{1}{N}$,
\begin{equation}\label{eqn:Hallhigher}
    Q^{(0)}=\frac{\xi}{2\pi}\Phi^{(1)},\quad \xi=-\frac{1}{N}~,
\end{equation}
where $Q^{(0)}$ is the charge of ordinary symmetry in a region, and $\Phi^{(1)}$ is the flux of the two-form gauge field for the one-form symmetry in the region. 
Similar transports of higher-form symmetry have been studied in hydrodynamics, see e.g. \cite{Das:2023nwl}.

Thus the models manifestly have nonzero generalized Hall conductivities, and also admit description in terms of local commuting projector models, i.e. toric codes.

\subsection{Protected gapless boundaries detect generalized Hall conductivities}

A signature of Hall conductivity is the gapless boundary modes, which are protected by anomalous $U(1)$ symmetries. 
The anomaly matches with the Hall conductivity in the bulk.
We can thus use the boundary theories as a detector to probe the generalized Hall conductivities in the bulk.
For the generalized Hall conductivity in (\ref{eq:generalhall}), we need to find a gapless boundaries with $U(1)_M\times U(1)_E$ symmetries that have a mixed anomaly. 
The boundary theories can be periodic scalar in (1+1)D, $U(1)$ gauge theory in (2+1)D and higher-form $U(1)$ gauge theories in higher dimensions.

However, it was known that gapless periodic scalar in (1+1)D or ordinary/higher-form $U(1)$ gauge theories in higher dimensions are challenging to realize on the lattice, due to the presence of monopole perturbation that will render the theory gapped without fine tuning. This is the generalization of the Polyakov's mechanism for confinement in $U(1)$ gauge theory in (2+1)D \cite{POLYAKOV1977429}. The lattice models such as \cite{Villain:1974ir,PhysRevB.16.1217} do not have both $U(1)\times U(1)$ symmetries, but only preserve one of the $U(1)$ due to the monopole perturbations. These lattice models are gapped without fine tuning and do not have the required anomalous symmetries to be the boundary theories.

Recently, these obstacles are overcome by the developments of the modified Villain formalism \cite{Gross:1990ub,Sulejmanpasic:2019ytl,Gorantla:2021svj} that constructs gapless ordinary/higher-form $U(1)$ gauge theories on the lattice with anomalous $U(1)\times U(1)$ symmetries by suppressing the monopole perturbations. In particular, the formalism produces lattice model for (1+1)D periodic scalar with both $U(1)$ winding and $U(1)$ momentum symmetries that have a mixed anomaly \cite{Cheng:2022sgb}. 

We will use the modified Villain formalism to construct symmetry-preserving gapless boundaries of the toric code models that preserve both $U(1)_M\times U(1)_E$ symmetries in various spacetime dimensions.
The fact that the models have such symmetry-preserving gapless boundaries shows that the local commuting projector models, i.e. toric codes with $U(1)$ symmetries, manifestly have nonzero generalized Hall conductivities.

\subsection{Symmetry junction operators on lattice}

\begin{figure}[t]
    \centering
    \includegraphics[width=0.8\linewidth]{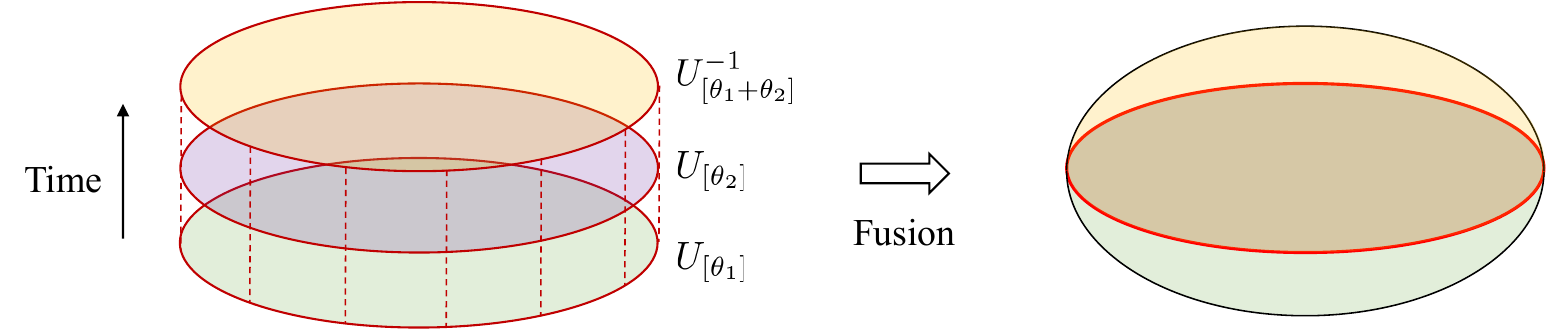}
    \caption{Symmetries $U_{[\theta_1]},U_{[\theta_2]},U_{[\theta_1+\theta_2]}$ meet at junction (red) in spacetime. The inverse for $U_{[\theta_1+\theta_2]}$ is due to relative orientation.}
    \label{fig:junctionopr}
\end{figure}

The reason that there can be nonzero Hall conductivities in commuting projector models and evade the no-go theorems lie in how symmetries are defined on the lattice. We will consider symmetries that do not act faithfully on the Hilbert space-- the symmetry operator $U(\Sigma)$ on closed oriented submanifold $\Sigma$ is trivial. The only nontrivial information of the symmetry is encoded in the junction of symmetry operators, and symmetry defects, i.e. excitation created by symmetry operator $U(\Sigma')$ with boundary $\partial\Sigma'\neq 0$. The symmetry defect gives a junction of symmetries with transformations $[\theta_1],[\theta_2],[\theta_1+\theta_2]$:
\begin{equation}
    \alpha_{\theta,\theta'}(\partial\Sigma'):=U_{[\theta_1+\theta_2]}(\Sigma')^{-1}U_{[\theta_2]}(\Sigma')U_{[\theta_1]}(\Sigma')~.
\end{equation}
The operator $\alpha_{\theta_1,\theta_2}$ is supported on $\partial\Sigma'$ since the right hand side composes to the identity in the absence of boundaries $\partial\Sigma'=0$. Since there is no net defects created by the operator on the right hand side, $\alpha_{\theta_1,\theta_2}(\partial\Sigma')$ is a symmetry operator and it encodes nontrivial information about the symmetry. We will call it symmetry junction operator.
This is the lattice counterpart of the nontrivial junction  (\ref{eqn:junctionsym}) in the continuum, where the junction in spacetime is $\partial\Sigma'$ (see Figure \ref{fig:junctionopr}).

More generally, such symmetries illustrate the following properties. Given a symmetry supported on closed submanifold $U(\Sigma_\text{closed})$ on the lattice, if it is a constant depth local unitary circuit, it can be truncated to define on open submanifolds with boundaries $U(\Sigma')$. Such operator creates symmetry defects on the boundary $\partial\Sigma'$.
Importantly, there can be different choices of such truncation with different symmetry defects, and they correspond to {\it different symmetries}, even when the symmetry operators on closed submanifolds are identical. In other words, they have {\it same symmetry operator but different symmetry defects}.
We will comment about the relation to literature in section \ref{sec:fractionalization}.

We will illustrate the discussion by explicitly constructing such a non-onsite truncation of $U(1)$ symmetries in $\mathbb{Z}_N$ toric codes, which cannot be written as a sum of onsite charge operators and carries Hall conductivity. We note that the symmetries are not onsite-- they cannot be expressed as the sum of onsite charge operators in commuting projector models --despite the fact that the $U(1)$ symmetries are not anomalous.
This is in contrast to the situation in (1+1)D where it is recently shown that any non-anomalous symmetries can be transformed into onsite symmetries \cite{Seifnashri:2025vhf,Kapustin:2025nju}. 
We also remark that since the $U(1)$ symmetries in our models do not have local charges in the bulk, there are no current operators in the bulk. Therefore one cannot use Kubo formula~\cite{niu1984quantised} to compute the Hall conductivities on the lattice. 
Instead, we compute the generalized Hall conductivities using several alternative methods: surface currents on gapless boundaries of the lattice model, flux insertion argument, and many body Chern number~\cite{NiuThoulessWu, AvronSeiler,Hastings:2013cop, bachmann2018quantization}.

The work is organized as follows. 
In section \ref{sec:TQFT} we discuss $\mathbb{Z}_N$ TQFT enriched with $U(1)$ symmetries and its gapless boundary in the continuum field theory language.
In section \ref{sec:ordinaryhall2d} we discuss $\mathbb{Z}_N$ toric code model in (2+1)D enriched with $U(1)$ symmetry with non-vanishing Hall conductivity.
In section \ref{sec:generalizedhall3d} we discuss $\mathbb{Z}_N$ toric code model in (3+1)D enriched with $U(1)$ ordinary and $U(1)$ one-form symmetries with non-vanishing generalized Hall conductivity.
In section \ref{sec:protectededge} we present arguments of protected gapless boundary modes for non-vanishing quantum Hall and generalized Hall conductivities.
In section \ref{sec:outlook} we comment on future directions.
We review the modified Villain formalism \cite{Gross:1990ub,Sulejmanpasic:2019ytl,Gorantla:2021svj} for constructing the gapless surface modes in Appendix \ref{app:modifiedVillain}.

\section{Topological Field Theory Enriched with $U(1)$ Symmetries}
\label{sec:TQFT}

Let us consider $\mathbb{Z}_N$ topological $q$-form gauge theory in $D$ spacetime dimension. The theory can be described by $q$-form gauge field $a$ and $(D-q-1)$ gauge field $b$, with the action \cite{Horowitz:1989ng,Kapustin:2014gua}
\begin{equation}\label{eqn:TQFT}
    \frac{N}{2\pi}\int adb~.
\end{equation}
The theory has $\mathbb{Z}_N$ operators $W_E=e^{i\oint a}$ and $W_M=e^{i\oint b}$. We will refer the the former as the electric operators and the later as the magnetic operators. These operators generate $\mathbb{Z}_N^{(D-q-1)}\times \mathbb{Z}_N^{(q)}$ higher-form symmetries, where we use superscript $(n)$ to denote an $n$-form symmetry.

We can enrich the system with $U(1)^{(D-q-2)}_E\times U(1)^{(q-1)}_M$ symmetries: when we turn on the corresponding background $U(1)$ gauge fields $A_E,A_M$, the action is modified to be
\begin{equation}
    \frac{N}{2\pi}\int adb+\frac{1}{2\pi}\int adA_E+\frac{1}{2\pi}\int A_Mdb~,
\end{equation}
where we consider the ``minimal coupling'' without additional classical topological terms for $A_E,A_M$. We note that the couplings are equivalent to activating background gauge fields for the $\mathbb{Z}_N^{(D-q-1)}\times \mathbb{Z}_N^{(q)}$ higher-form symmetries in terms of $A_E,A_M$ \cite{Benini:2018reh,Hsin:2019fhf}.
The corresponding conserved currents are
\begin{equation}
    J_E=\frac{1}{2\pi}(-1)^{(q+1)D}\star da,\quad J_M=\frac{1}{2\pi}\star db~,
    \label{eq:u1currents}
\end{equation}
where we used $\frac{1}{2\pi}\int adA_E=(-1)^{q+1}\frac{1}{2\pi}\int da A_E=(-1)^{(q+1)D}\frac{1}{2\pi}\int A_Eda$.
The equation of motion for $a,b$ indicates that
\begin{equation}
    b=-\frac{1}{N}A_E,\quad a=-\frac{1}{N}A_M~.
\end{equation}
Thus the electric operator $W_E$ carries fractional charge $(-1/N)$ under the $U(1)_M^{(q-1)}$ symmetry, while the magnetic operator $W_M$ carries fractional charge $(-1/N)$ under the $U(1)^{(D-q-2)}$ symmetry. 

The currents satisfy
\begin{equation}
    J_E=(-1)^{(q+1)D}\frac{\xi}{2\pi}\star dA_M,\quad 
    J_M=\frac{\xi}{2\pi}\star dA_E,\quad \xi=-\frac{1}{N}~.
\end{equation}
Such transport equation can be represented by a mixed Chern-Simons like term $\frac{\xi}{2\pi}A_MdA_E$.

\subsection{Gapless boundary state}

A gapless boundary state that preserves the symmetries is $U(1)$ $(q-1)$-form gauge theory in $(D-1)$ spacetime dimension. 
The theory is gapless, and conformal if $D=2q$.
Denote the $(q-1)$-form gauge field by $u$, the theory has $U(1)^{(q-1)}\times U(1)^{(D-q-2)}$ symmetry, with conserved currents ($e$ is the gauge coupling) \cite{Gaiotto:2014kfa}
\begin{equation}
    j_M=\frac{1}{e^2}du,\quad j_E=\star \frac{1}{2\pi}du~.
\end{equation}
The symmetries have a mixed 't Hooft anomaly: if we turn on background gauge fields $B_M,B_E$ for the symmetries (the notation is to match with the bulk gauge fields $A_E,A_M$), the action is
\begin{equation}
    -\frac{1}{2e^2}\int (du-B_M)\star (du-B_M)+\frac{1}{2\pi}\int du B_E~.
\end{equation}
The theory is not invariant under the background gauge transformation. The anomaly is described by the bulk term
\begin{equation}
    \frac{1}{2\pi}B_MdB_E~.
\end{equation}
Thus we can put the theory on the boundary of the TQFT (\ref{eqn:TQFT}) provided that the $U(1)$ symmetry is extended to $N$-fold covering: for example, if we couple the boundary theory to the bulk with the boundary conditions $a|=du=2\pi\star j_E $, $b|=\frac{2\pi}{Ne^2}\star du=\frac{2\pi}{N} \star j_M$,
\begin{equation}
    NB_M=A_M|,\quad B_E=A_E|~.
\end{equation}
Example of the gapless boundary state in higher dimension $D=5,q=2$ is discussed in e.g. \cite{Chen:2021xuc}.

\section{Toric Code with Nonzero $U(1)$ Hall Conductivity}
\label{sec:ordinaryhall2d}

\begin{figure}[h]
    \centering
    \includegraphics[width=0.45\textwidth]{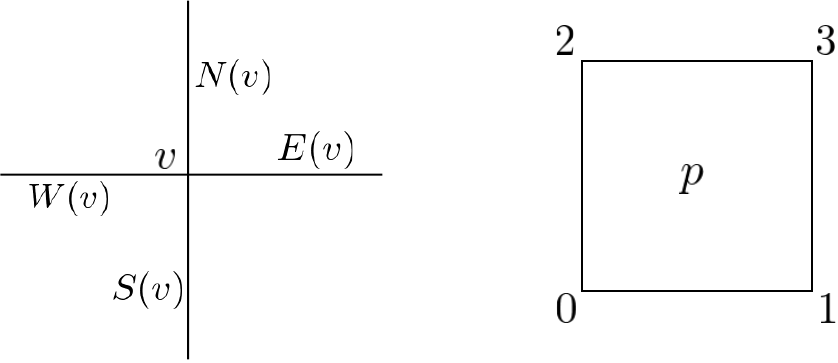}
    \caption{The edges nearby a vertex $v$ and plaquette $p$.}
    \label{fig:doublemodel}
\end{figure}

Let us consider $\Z_N$ toric code with $U(1)$ symmetry in (2+1)D. 
We consider a square lattice with a $\Z_N$ qudit on each edge, with the Hamiltonian
\begin{align}
    H_{\Z_N} =  -\sum_v A_v - \sum_p B_p + \text{h.c.}~,
\end{align}
where $v$ and $p$ denote the vertex and plaquette respectively. Specifically, each term is given by
\begin{align}
    A_v = {X}_{N(v)}{X}_{E(v)}{X}^\dagger_{W(v)}{X}^\dagger_{S(v)}~,
\end{align}
\begin{align}
    B_p = Z_{01}Z_{13}Z^\dagger_{02}Z^\dagger_{23}~,
\end{align}
with $\Z_N$ qudit Pauli operators satisfying $ZX=e^{2\pi i/N} XZ$.
These two terms are illustrated in Figure \ref{fig:doublemodel}.

We define the $\Z_N$ gauge field $a$ on links satisfying $e^{\frac{2\pi i}{N}a_e} = Z_e$. The state with $B_p=1$ corresponds to the flat $\Z_N$ gauge field $da=0$. $A_v=1$ corresponds to the Gauss law constraint of the $\Z_N$ gauge field, so the ground state Hilbert space is described by the $\Z_N$ gauge theory. Similarly, we also define a dual $\Z_N$ gauge field $b$ on dual links $\hat e$ satisfying $e^{\frac{2\pi i}{N} b_{\hat e}}=X_{\hat e}$.
The state with $A_v=1$ corresponds to the flat $\Z_N$ gauge field $db=0$, and $B_p=1$ enforces the Gauss law for the dual gauge field $b$.

The theory has electric and magnetic string operators that generate $\mathbb{Z}_N\times\mathbb{Z}_N$ one-form symmetry:
\begin{equation}
    W_e(\gamma)=e^{\frac{2\pi i}{N}\oint_\gamma a},\quad  W_m(\tilde \gamma)=e^{\frac{2\pi i}{N}\oint_{\tilde\gamma} b}~,
\end{equation}
where $\gamma$ is a closed loop on the lattice and $\tilde\gamma$ is a closed loop on the dual lattice.

\paragraph{Ordinary $U(1)_E\times U(1)_M$ symmetry}
One can define two ordinary $U(1)$ symmetry operators $U^e_{\theta},U^m_{\theta}$ as the following action on the state
\begin{align}\label{eqn:U(1)symmetry}
    U^e_{\theta} = \exp\left(i[\theta] \int \frac{d\tilde{a}}{N}\right), \quad 
    U^m_{\theta} = \exp\left(i[\theta] \int \frac{d\tilde{b}}{N}\right).
\end{align}
where $\tilde a$ is the integral lift of $a$ to $\{0,\dots,N-1\}\in\Z$, and $\tilde b$ is the same integral lift of $b$. $[\theta]$ denotes the restriction to $[0,2\pi)$. 
The operator $d\tilde{a}$ is defined on each plaquette, which is given by
\begin{align}
    d\tilde{a}(0123) = \tilde a(01) + \tilde a(13) - \tilde a(02) - \tilde a(23)
\end{align}
with $\tilde{a} = \mathrm{diag}(0,1,\dots, N-1)$ in the $Z$ eigenbasis.
We note that each circuit $\exp(i[\theta]d\tilde a/N)$ on a plaquette becomes product of a fractional power of Pauli $Z$ operators,
\begin{align}
    \exp\left(i[\theta] \frac{d\tilde{a}(0123)}{N}\right) = Z^{\frac{[\theta]}{2\pi}}_{01} Z^{\frac{[\theta]}{2\pi}}_{13} Z^{\frac{[\theta]}{2\pi}\dagger}_{02} Z^{\frac{[\theta]}{2\pi}\dagger}_{23}~.
\end{align}
The integral then represents the sum of  $d\tilde{a}$ operators over the plaquettes, so we have
\begin{align}
    U_\theta^e =  \exp\left(i[\theta] \int \frac{d\tilde{a}}{N}\right) = \prod_{p=(0123)}Z^{\frac{[\theta]}{2\pi}}_{01} Z^{\frac{[\theta]}{2\pi}}_{13} Z^{\frac{[\theta]}{2\pi}\dagger}_{02} Z^{\frac{[\theta]}{2\pi}\dagger}_{23}~.
\end{align}
Similarly, the operator $d\tilde b$ is defined on a dual plaquette. Each dual gauge field $\tilde b$ on the dual link is given by
\begin{align}
    \tilde b(\hat e) = H \tilde a H^{\dagger}(\hat e),
\end{align}
with $H$ the Hadamard like operator with the matrix elements
\begin{align}
    H_{jk} = \frac{1}{\sqrt{N}}\exp(\frac{2\pi i}{N} jk)~, 
\end{align}
so each circuit $\exp(i[\theta]d\tilde b/N)$ on a dual plaquette (labeled by a vertex $v$) becomes
\begin{align}
    \exp\left(i[\theta] \frac{d\tilde{b}(v)}{N}\right) = X^{\frac{[\theta]}{2\pi}}_{N(v)} X^{\frac{[\theta]}{2\pi}}_{E(v)} X^{\frac{[\theta]}{2\pi}\dagger}_{W(v)} X^{\frac{[\theta]}{2\pi}\dagger}_{S(v)}~,
\end{align}
and
\begin{align}
    U_\theta^m = \exp\left(i[\theta]\int \frac{d\tilde{b}}{N}\right) = \prod_v X^{\frac{[\theta]}{2\pi}}_{N(v)} X^{\frac{[\theta]}{2\pi}}_{E(v)} X^{\frac{[\theta]}{2\pi}\dagger}_{W(v)} X^{\frac{[\theta]}{2\pi}\dagger}_{S(v)}~.
\end{align}
These symmetry operators $U_\theta^e,U_\theta^m$ are lattice counterpart of $U(1)_E\times U(1)_M$ symmetries in \eqref{eq:u1currents}.
The $U(1)$ symmetry with nonzero Hall conductance is given by the diagonal element in $U(1)_E\times U(1)_M$ symmetry,
\begin{align}
    U^{em}_\theta =\exp\left[i[\theta] \left(\int \frac{d\tilde{b}}{N}\right)\right]\exp\left[i[\theta] \left(\int \frac{d\tilde{a}}{N}\right)\right]~.
    \label{eq:Uemdef}
\end{align}
This expression implies that $U^{em}_\theta$ is a depth 2 circuit where local unitaries are given by evaluating $d\tilde a$ or $d\tilde b$ on a small plaquette (or that on a dual lattice).

\paragraph{Symmetry condition}

The operators $U^e_\theta,U^m_\theta$ are symmetries since they are total derivative: they commute with Hamiltonian terms in the bulk. Moreover, they give rise to symmetry defects, which are fractional power $e$ and $m$ strings.
We also need to check the junction of symmetry operators commute with Hamiltonian, e.g. $U_{\theta_1}U_{\theta_2}U_{[\theta_1+\theta_2]}^{-1}$ with $[\cdot]$ denote the restriction to $[0,2\pi)$ 
for any finite region does not create any symmetry defects, and it should commute with the Hamiltonian: indeed the combination is an integer power of the $e$ string or $m$ string for $U_\theta=U^e_\theta$ and $U_\theta=U^m_\theta$, respectively.

\subsection{Protected gapless edge states with $U_{\theta}^{em}$ symmetry}
\label{subsec:1dmV}

Here let us present a gapless boundary theory of the (2+1)D $\Z_N$ toric code that preserves the $U(1)$ symmetry $U_\theta^{em}$ with nonzero Hall conductance. The boundary theory is realized by a modified Villain model \cite{Gross:1990ub,Sulejmanpasic:2019ytl,Gorantla:2021svj} in (1+1)D, which describes a (non-chiral) $c=1$ free compact boson.

\paragraph{Review of modified Villain model in (1+1)D}
First, let us briefly review the modified Villain model in (1+1)D without coupling to the bulk, following \cite{Cheng:2022sgb}. The idea is that we start with a lattice Hamiltonian with a non-compact boson with $\mathbb{R}$ symmetry, and then gauge $\Z$ subgroup to get a compact boson with $U(1)$ symmetry. Namely, we first consider a matter Hamiltonian
\begin{align}
    H_{\text{matter}} = \sum_j \left( \frac{U_0}{2}p_j^2 + \frac{J_0}{2}(\phi_{j+1}-\phi_j)^2 \right)~,
\end{align}
with
\begin{align}
    [\phi_j,p_{j'}] = i\delta_{j,j'}~.
\end{align}
The matter $\phi_j$ is $\mathbb{R}$-valued, and the model has the $\mathbb{R}$ shift symmetry $\phi_j\to \phi_j+\xi$. Its subgroup $\mathbb{Z}\subset \mathbb{R}$ is then gauged by introducing the link $\Z$ variables $n_{j,j+1}$, which yields the Hamiltonian of the modified Villain model
\begin{align}
    H_{\text{mV}} = \sum_j \left( \frac{U_0}{2}p_j^2 + \frac{J_0}{2}(\phi_{j+1}-\phi_j-2\pi n_{j,j+1})^2 \right)~,
\end{align}
with 
\begin{align}
    [n_{j,j+1}, E_{j',j'+1}] = i\delta_{j,j'}~,
\end{align}
which is subject to the Gauss law constraint
\begin{align}
    G_j = e^{i(E_{j,j+1}-E_{j-1,j})}e^{-2\pi i p_j} = 1~.
\end{align}
This modified Villain model has the $U(1)_m\times U(1)_w$ symmetry whose generators are given by
\begin{align}
    Q_m = \sum_j q_j^m, \quad q_j^m = p_j~,
\end{align}
\begin{align}
    Q_w = \sum_j q_{j,j+1}^w, \quad q_{j,j+1}^w = \frac{1}{2\pi}(\phi_{j+1}-\phi_j-2\pi n_{j,j+1})~,
\end{align}
where the periodicity of $U(1)_w$ follows from the Gauss law, $\exp(2\pi iQ_m) = \prod_j G_j =1$.
The $U(1)_m\times U(1)_w$ symmetry has the mixed 't Hooft anomaly
\begin{align}
    \frac{1}{2\pi}\int A^{(m)}dA^{(w)}~.
\end{align}

\begin{figure}[t]
    \centering
    \includegraphics[width=0.55\textwidth]{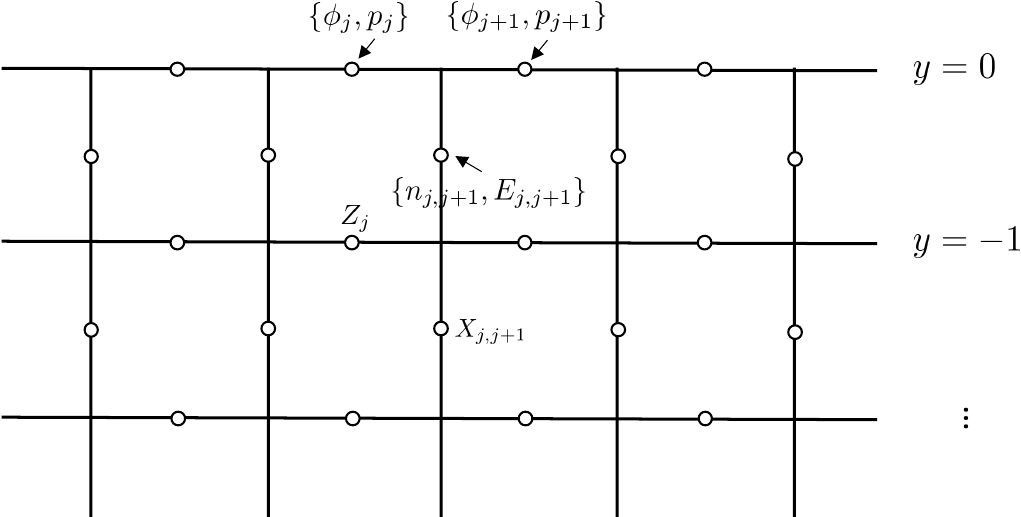}
    \caption{The smooth boundary of the square lattice is located at $y=0$, and the square lattice is identified as a Euclidean lattice. The degrees of freedom nearby the boundary ($y>-1$) are replaced with the Villain fields. The qudits are at edges with $y\le -1$.}
    \label{fig:TCboundary}
\end{figure}

\begin{figure}[t]
    \centering
    \includegraphics[width=0.8\textwidth]{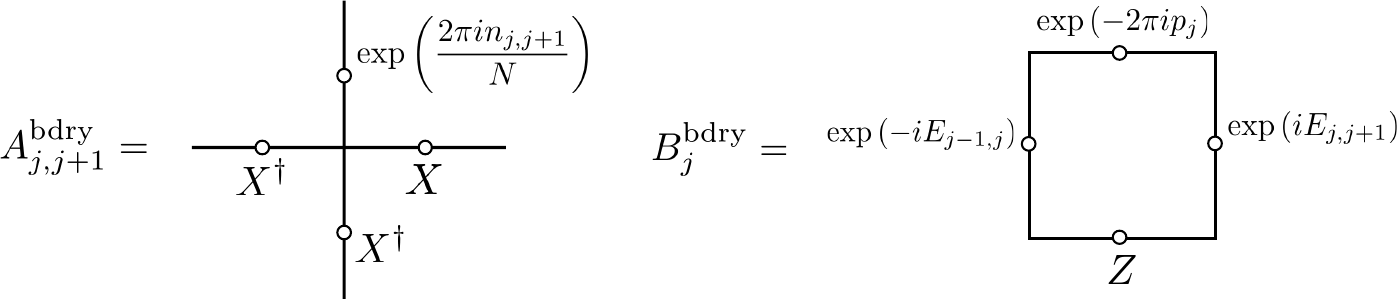}
    \caption{The definitions of the boundary Hamiltonian near the boundary.}
    \label{fig:Hamboundary}
\end{figure}

\paragraph{Coupling to the bulk} 
 Now let us couple the above modified Villain model to the (2+1)D $\Z_N$ toric code in the bulk. We consider the square lattice with smooth boundary, and replace the $\Z_N$ qudits with the Villain fields near the boundary as shown in Figure \ref{fig:TCboundary}.
\begin{align}
    H= -\sum_{v\subset \text{bulk}} A_v - \sum_{p\subset\text{bulk}} B_p + 
    H_{\text{mV}}^{\text{bdry}} - g\left(\sum_jA^{\text{bdry}}_{j,j+1} +\sum_{j} B^{\text{bdry}}_j\right) + \text{h.c.}
\end{align}
where $j$ labels the position at the boundary, and the definitions of the boundary Hamiltonian is found in Figure \ref{fig:Hamboundary}. $H_{\text{mV}}^{\text{bdry}}$ is the modified Villain Hamiltonian using the boundary Villain fields. We do not enforce the Gauss law $G_j=1$ at the boundary, but the boundary stabilizer $B^{\text{bdry}}_j=1$ is thought of as a modified Gauss law coupled to the bulk qudit, which we energetically enforce by $g\gg 1$.

Let us study the symmetry of the bulk-boundary system. We again consider the $U(1)$ symmetry $U^{em}$ in this bulk-boundary Hamiltonian. The operator $U_{\theta}^{em}$ is defined by replacing the local circuit $\exp(i[\theta]d\tilde a/N )$ in \eqref{eq:Uemdef} at the boundary by
\begin{align}
    \exp\left(i[\theta]\frac{d\tilde a_{0123}}{N} \right) \to \exp\left(i[\theta]\left(\frac{\tilde a_{01}}{N} + \frac{E_{13}}{2\pi} - \frac{E_{02}}{2\pi} - p_{23}\right)\right)~,
\end{align}
and replacing $\exp(i[\theta]d\tilde b/N )$ at the boundary by
\begin{align}
    \exp\left(i[\theta]\frac{d\tilde b_{v}}{N} \right) \to \exp\left(i[\theta]\left(\frac{\tilde b_{E(v)}}{N} - \frac{\tilde b_{W(v)}}{N}+ \frac{n_{N(v)}}{N} -\frac{\tilde b_{S(v)}}{N}\right)\right)~,
\end{align}
where the labels of edges follow Figure \ref{fig:doublemodel}. 

When the $U(1)$ symmetry $U_{\theta}^{em}$ acts on the whole system, the symmetry acts solely on the boundary Villain fields;
\begin{align}
    U_{\theta}^{em} = \prod_j \exp(-i\theta p_j) \prod_j \exp\left(\frac{i\theta}{N}n_{j,j+1}\right)
    \label{eq:boundary U(1) action}
\end{align}
This corresponds to $(\eta_m,\eta_w)=\left(-\theta,\frac{\theta}{N}\right)\in U(1)_m\times U(1)_w$.

\paragraph{Symmetry extension on the boundary}

We note that due to the identification $\eta_w=\frac{\theta}{N}$ for $2\pi$-periodic $\eta_w$, $\theta$ is no longer $2\pi$ periodic on the boundary: it becomes $2\pi N$ periodic. 
Within the low energy Hilbert space $B_j^{\text{bdry}}=1$, the symmetry at the boundary $U(1)_m$ is extended by $\Z_N$;
\begin{align}
    \exp(2\pi iQ_m) = \prod_{j} B_j^{\dagger \text{bdry}} Z_j = \prod_j Z_j~,
\end{align}
i.e., the momentum $U(1)_m$ symmetry with $\eta_m=2\pi$ is identified as the $e$ Wilson line operator parallel to the boundary. Therefore the symmetry is extended as
\begin{align}
    \Z_N\to \tilde U(1)_m\to U(1)_m~.
\end{align}
The winding $U(1)_w$ symmetry is not extended, but $\eta_w=2\pi/N$ is identified as the line operator for $m$ particle within the low-energy Hilbert space $A_{j,j+1}^{\text{bdry}}=1$,
\begin{align}
    \exp\left(\frac{2\pi i}{N} Q_w\right) = \exp\left(\frac{2\pi i}{N} \sum_j n_{j,j+1}\right) = \prod_j A_{j,j+1}^{\text{bdry}} X_{j,j+1} = \prod_j X_{j,j+1}~.
\end{align}
It then follows that the $U_\theta^{em}$ with $\theta=2\pi$ is identified as the $em$ line operator parallel to the boundary,
\begin{align}
    U_{2\pi}^{em} = \prod_j X_{j,j+1} \prod_j Z^\dagger_j~,
\end{align}
which means the group structure at the boundary is given by the symmetry extension
\begin{align}
    \Z_N^{em}\to \tilde U(1)^{em}\to U(1)^{em}~.
\end{align}

\paragraph{Anomaly of the boundary theory} 

In the bulk, the $U(1)$ symmetry acts on the theory through symmetry fractionalization with the $\Z^{em}_N$ 1-form symmetry. In particular, the $U(1)_E\times U(1)_M$ 0-form symmetry generated by $U^e_\theta,U^m_\theta$ has a mixed response: the low energy subspace of the bulk van be described by $\mathbb{Z}_N$ gauge theory with $U(1)$ gauge field $a$ and $U(1)$ dual gauge field $b$: 
\begin{equation}
    \frac{N}{2\pi} adb+\frac{1}{2\pi}adA_E +\frac{1}{2\pi}bdA_M~,
\end{equation}
where $A_E,A_M$ are the background gauge fields for the $U(1)$ 0-form symmetries, and the coupling follows from (\ref{eqn:U(1)symmetry}). Integrate out $a,b$ gives the fractional response: denote the $U(1)$ currents by $J_E=\star da/(2\pi),J_M=\star db/(2\pi)$ with Hodge star operation $\star$, it is related to the background magnetic fields as
\begin{equation}
    J_E=-\frac{(1/N)}{2\pi} \star dA_M,\quad  J_M=-\frac{(1/N)}{2\pi} \star dA_E~.
\end{equation}
The transport can be represented by a mixed Chern-Simons term between $A_E,A_M$ of effective level $-1/N$.

On the boundary, the $U(1)_M$ symmetry is $2\pi N$ periodic and extended to be $N$-fold covering, which rescales $A_M\rightarrow N\tilde A_M$. 
The response of $U(1)_E\times U(1)_M$ becomes a properly quantized mixed Chern-Simons term for $U(1)_E\times \tilde U(1)_M$ with level $N\times (-1/N)=-1$.
Thus the boundary has $U(1)_E\times \tilde U(1)_M$ symmetry with mixed anomaly given by the mixed Chern-Simons term
\begin{equation}
-    \frac{1}{2\pi}A_E d\tilde A_M~.
\end{equation}
This is consistent with the mixed anomaly between the $U(1)$ winding and $U(1)$ momentum symmetry in the compact boson theory in (1+1)D.

\subsection{Non-vanishing Hall conductance}
\label{subsec:hall}

\subsubsection{Hall conductance from edge currents}

The Hall conductance can be computed from the commutator of charge density operators on the boundary. In our case, the charge densities of the $U(1)_m\times U(1)_w$ symmetries are $q_j^m=p_j,q_{j,j+1}^w=\frac{1}{2\pi}(\phi_{j+1}-\phi_j-2\pi n_{j,j+1})$. Using the canonical commutator $[\phi_j,p_{j'}]=i\delta_{j,j'}$ we find that \cite{Cheng:2022sgb}
\begin{equation}
    [q_{j,j+1}^w,q_{j'}^m]=\frac{i}{2\pi}\left(\delta_{j+1,j'}-\delta_{j,j'}\right)~.
\end{equation}
Thus the two $U(1)$ symmetries have a mixed anomaly with unit coefficient.\footnote{
This is the discrete analogue of $[\rho(x),\rho(y)]=\frac{i\sigma}{2\pi}\delta'(y-x)$.
} In terms of the bulk $U(1)$ symmetries related by $N$-fold covering, this indicates the mixed Hall conductivity $1/N$ between $U(1)_E\times U(1)_M$.

\subsubsection{Many-body Chern number}
\label{subsec:chernnumber}

\begin{figure}[t]
    \centering
    \includegraphics[width=0.7\textwidth]{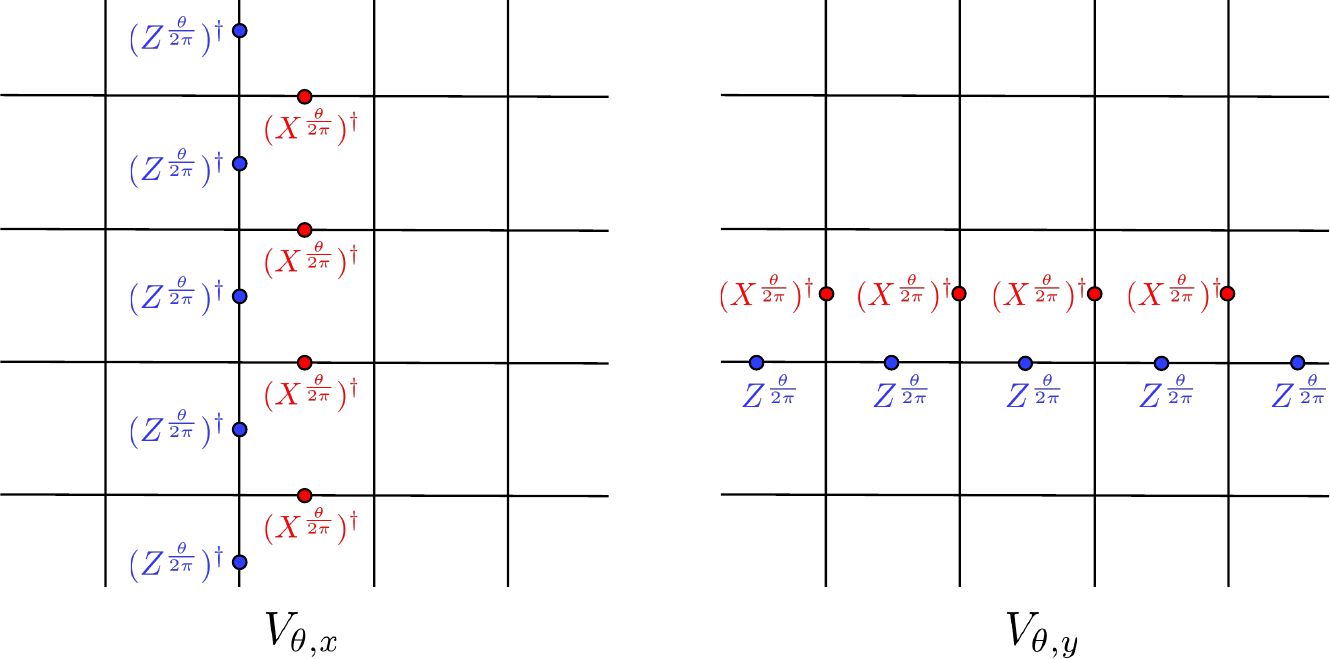}
    \caption{The operators $V_{\theta,x},V_{\theta,y}$ that introduces the twisted boundary condition. This is regarded as the $U(1)$ symmetry defect for $U^{em}_{\theta}$ with our choice of truncation.}
    \label{fig:twist}
\end{figure}

One way to define Hall conductance with the diagonal $U(1)$ symmetry is to use the many-body Chern number~\cite{NiuThoulessWu, AvronSeiler}. This allows us to compute the Hall conductance purely from a bulk, without introducing boundaries. Let us consider the system on a torus $S^1_x\times S^1_y$, with twisted boundary condition for $U(1)$ symmetry $\{\theta_x,\theta_y\}$. The twisted boundary condition is described by a defect Hamiltonian
\begin{align}
    H[\theta_x,\theta_y] = V_{\theta,x}  V_{\theta,y} H[0,0] V^\dagger_{\theta,y} V^\dagger_{\theta,x}
\label{eq:defectham}
\end{align}
with the untwisted Hamiltonian $H[0,0]$ without defects,
where $V_\theta$ is a line operator that generates the $U(1)$ symmetry defect $U^{em}_\theta$ according to our truncation of the symmetry operators. See Figure \ref{fig:twist}.
The operator $V_\theta$ is identified as the symmetry operator for the anyon $em$ and commutes with the Hamiltonian. Therefore we have $H[0,0]=H[0,2\pi] = H[2\pi,0]=H[2\pi,2\pi]$. $H[\theta_x,\theta_y]$ with generic twist $(\theta_x,\theta_y)$ is a commuting projector Hamiltonian with gapped ground states.

The many-body Chern number is then characterized by integrating Berry curvature of ground states over the parameter space $T^2$ for the twist $(\theta_x,\theta_y)$. 
This can be computed by evaluating the Berry phase along the adiabatic deformation of a fixed ground state $\ket{\Psi[0,0]}$ without twist, through a path~\cite{Hastings2014quantization}
\begin{align}
    \ket{\Psi[0,0]}\to \ket{\Psi[2\pi,0]}\to \ket{\Psi[2\pi,2\pi]}\to \ket{\Psi[0,2\pi]}\to \ket{\Psi[0,0]}~,
\end{align}
which goes around the whole area of a torus. Each adiabatic insertion of flux is realized by the operator $V_{2\pi}$ or its inverse, for instance $\ket{\Psi[2\pi,0]} = V_{2\pi,y}\ket{\Psi[0,0]}$.
Therefore the many-body Chern number $C$ becomes
\begin{align}
    \exp(2\pi i C) = \bra{\Psi[0,0]}V^\dagger_{2\pi,x}V^\dagger_{2\pi,y}V_{2\pi,x}V_{2\pi,y}\ket{\Psi[0,0]} =e^{4\pi i/N}~,
\end{align}
which shows that the many-body Chern number becomes $2/N$ mod 1.

\subsubsection{Flux insertion}

\begin{figure}[t]
    \centering
    \includegraphics[width=0.6\linewidth]{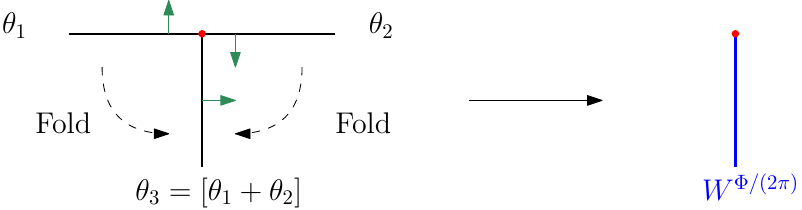}
    \caption{Junction of symmetry defects, where green sold arrows at each line indicates the orientation of the defects. After folded, the junction becomes the string operator $W^{\Phi/(2\pi)}$ with $\Phi=\theta_1+\theta_2-[\theta_1+\theta_2]$.}
    \label{fig:junction}
\end{figure}

The Hall conductance can be computed locally by the definition $Q_m=\frac{\sigma_{em}}{2\pi} \Phi_e$ for charge $Q_m$ for $U(1)_m$ carried by magnetic flux $\Phi_e$ for $U(1)_e$. Consider a junction of three symmetry defects for $U(1)_e$ symmetry on sphere space, with transformation parameters $\theta_1,\theta_2,\theta_3=[\theta_1+\theta_2]$ (see Figure \ref{fig:junction}). 
The magnetic flux at the center of the junction is
\begin{equation}
    \Phi_e=\theta_1+\theta_2-[\theta_1+\theta_2]\in 2\pi\mathbb{Z}~.
\end{equation}

We want to compute the $U(1)_m$ charge at the junction center, which is given by the value of the $Q_m=\int d\tilde b/N$ for any region enclosing the junction center. We only care about fractional part of the charge, so we want to compute the value of $e^{2\pi i Q_m}$, which is the value of unit $m$ string surrounding the junction center. In order to compute the charge at the center of the junction, we fold the symmetry defects as in Figure \ref{fig:junction}, then the only excitation is at the center of the folded junction, created by the string operator
\begin{equation}
    W_e^{\left(\theta_1+\theta_2-[\theta_1+\theta_2]\right)/2\pi}=W_e^{\Phi_e/(2\pi)}~,
\end{equation}
where the minus sign is due to the relative orientation when the defects are folded.
This gives the braiding phase $e^{\frac{2\pi i}{N}\left(\theta_1+\theta_2-[\theta_1+\theta_2]\right)/(2\pi)}=e^{\frac{i}{N}\left(\theta_1+\theta_2-[\theta_1+\theta_2]\right)}$, and thus
\begin{equation}
    Q_m=\frac{1}{2\pi N}\left(\theta_1+\theta_2-[\theta_1+\theta_2]\right)=\frac{1}{N}\frac{\Phi_e}{2\pi} \text{ mod }1~.
\end{equation}
Thus we conclude that the fractional Hall transport between $U(1)_e,U(1)_m$ is
\begin{equation}
    \sigma_{em}=\frac{1}{N}\text{ mod }1~.
\end{equation}
The transport represents the mixed Chern-Simons response $(\sigma_{em}/2\pi) A_edA_m$.

The Hall transport for the diagonal $U(1)\subset U(1)_e\times U(1)_m$ symmetry can be computed in a similar way. By setting $A_e=A_m=A$, this gives fractional Hall response $\sigma=2/N$ mod 2 with effective Chern-Simons term $(\sigma/4\pi)AdA$.

\subsection{Evasion of no-go theorems for vanishing Hall conductivity}

In the no-go theorems in \cite{Kapustin_2019,Zhang2022hall} for vanishing Hall conductivity for local commuting projector model with onsite $U(1)$ symmetry, it is assumed that the $U(1)$ charge is a sum of local terms $Q(\Lambda)=\sum_{\lambda\in \Lambda }Q_\lambda $ with well-defined local charge $Q_\lambda$ with integer eigenvalues on each local Hilbert space ${\cal H}_\lambda$, ${\cal H}=\otimes_\lambda {\cal H}_\lambda$. In other words, each local Hilbert space is acted on by charge operator $Q_\lambda$.

On the other hand, for our $U(1)$ symmetry, the charge for a region $\Lambda$ is localized on the boundary $\partial\Lambda$: the charge acts trivially on local Hilbert space ${\cal H}_\lambda$ inside the region away from the boundary. The charge operator is thus pertinent to a chosen region.

Let us show that there does not exist well-defined local charge $Q_i$ such that for all region $R$,
\begin{equation}
    \text{(Does not exist)}\quad \int_R d\tilde a=\sum_{i\in R} Q_i~.
\end{equation} 
To see this, suppose the relation were true. Then  $Q_i=0$ for any site in the interior of region $R$. We can cover the space with overlapping regions $\{R_I\}$ such that each site is in the interior of some region $R_I$. Thus we conclude that $Q_i=0$ for all sites. On the other hand, the charge operator $\int_R d\tilde a$ can have nonzero eigenvalues, and thus we have a contradiction.

Similarly, there is also no canonical current $J_{j,j'}=i[H_{j'},Q_{j}]-i[H_{j},Q_{j'}]$ where $H_j$ are the local Hamiltonian terms $H=\sum H_j$ due to the absence of $Q_j$.

We note that even in the absence of a well-defined local charge operator, one can still define twisted boundary conditions and corresponding defect Hamiltonians $H[\theta_x, \theta_y]$ by introducing $U(1)$ symmetry defects constructed via a specific truncation procedure (see, e.g., Eq.\eqref{eq:defectham}). In general, such $H[\theta_x, \theta_y]$ is not periodic (i.e., not a trigonometric function) in $\theta_x$ and $\theta_y$. For instance, the Hamiltonian in Eq.\eqref{eq:defectham} has $H[\theta_x, \theta_y] \neq H[\theta_x + 2\pi, \theta_y]$, except when $\theta_y = 0 \mod 2\pi$. In contrast, when a well-defined local integral charge exists, $H[\theta_x, \theta_y]$ (with canonical choice of defects from onsite truncation) becomes a trigonometric function of the twists. In such cases, as shown in Ref.~\cite{Kapustin_2019}, any commuting projector Hamiltonian with finite dimensional local Hilbert spaces necessarily has vanishing many-body Chern number, and hence zero Hall conductance.

This distinction underlies our mechanism for evading the no-go theorems of Refs.~\cite{Kapustin_2019,Zhang2022hall}, thereby allowing for nonzero Hall conductivity in local commuting projector models.

\subsection{Comments on Fractionalization}
\label{sec:fractionalization}

Both of these $U(1)$ symmetry exhibit the symmetry fractionalization \cite{Barkeshli:2014cna,Teo_2015,Tarantino_2016,Benini:2018reh}, with a vison $v=e,m$ respectively. To see this for $U_\theta^e$, we define the symmetry generator within a disk region $R$ that consists of a collection of plaquettes. The fusion rule of the operators is then modified by the closed line operator along the circle $\partial R$,
\begin{align}
    U^e_\theta (R) \times U^e_{\theta'} (R) = U^e_{\theta+\theta'} (R) \times W_e(\partial R)^{\eta(\theta,\theta')}~,
\end{align}
with $\eta(\theta,\theta') = (\theta+\theta'-[\theta+\theta'])/2\pi$.
This implies symmetry fractionalization with the vison $v=e$. Similarly, one can define the $U_\theta^m$ on the disk region $\hat{R}$ that consists of a collection of dual plaquettes. Then
\begin{align}
    U^m_\theta (\hat R) \times U^m_{\theta'} (\hat R) = U^m_{\theta+\theta'} (\hat R) \times W_m(\partial \hat R)^{\eta(\theta,\theta')}~.
\end{align}

Similarly, the symmetry operator $U^{em}_\theta$ with diagonal $U(1)_E\times U(1)_M$ action can be defined on a disk region $(R,\hat{R})$, where $R$ is a disk region given by a collection of plaquettes, and $\hat{R}$ is obtained by the largest region formed by plaquettes of the dual lattice inside $R$,
\begin{align}
    U^{em}_\theta(R,\hat{R}) =\exp\left[i[\theta] \left(\int_{\hat R} \frac{d\tilde{b}}{N}\right)\right]\exp\left[i[\theta]\left(\int_R \frac{d\tilde{a}}{N}\right)\right]=\exp\left[i[\theta]\left(\int_R \frac{d\tilde{a}}{N} +\int_{\hat R} \frac{d\tilde{b}}{N}\right)\right].
\end{align}
where the last expression follows since the two layers of the circuit are commutative after the integral.

This operator exhibits the symmetry fractionalization
\begin{align}
     U^{em}_\theta (R,\hat R) \times U^{em}_{\theta'} (R, \hat R) = U^{em}_{\theta+\theta'} (R,\hat R) \times W_e(\partial R)^{\eta(\theta,\theta')}W_m(\partial \hat R)^{\eta(\theta,\theta')}~.
\end{align}
This $U(1)$ symmetry has a vison $v=em$ carrying the spin $1/N$, hence leads to the Hall conductance $2/N$. Indeed, its $\theta=2\pi$ kink (domain wall) at the 2d space is realized by a line operator for $em$ along the domain wall, reflecting that the vison is given by $em$. In Sec.~\ref{subsec:hall} we compute the many-body Chern number to demonstrate the Hall conductance.

Let us note special points about the above symmetry operators:
\begin{itemize}
    \item $U^e_\theta, U^m_\theta, U^{em}_\theta$ are given by local finite depth unitary, where each local unitary is given by $d\tilde a$ or $d\tilde b$ on a plaquette (or a dual plaquette).
    \item $U^e_\theta, U^m_\theta, U^{em}_\theta$ are identity operators when defined on a closed oriented 2d space.\footnote{One can make $U(1)$ symmetry act faithfully on the lattice model, by stacking with a trivial gapped state with onsite $U(1)$ symmetry and consider the diagonal $U(1)$ action on the trivial and the toric code state. This clearly does not change the response of the $U(1)$ symmetry.} 
    Hence, these symmetry operators does not act faithfully on a closed 2d space. They only act in a nontrivial way on a 2d space with a boundary, in which case the symmetry operators act within the boundary. For instance, when the 2d space $\Sigma$ has a boundary, $U_e^\theta$ acts by
    \begin{align}
        U_\theta^e(\Sigma) = \exp\left(\frac{i[\theta]}{N}\int_{\partial\Sigma} \tilde{a}\right)
    \end{align}
    \item Though the symmetry operator on a closed space is trivial, the symmetry defect (domain wall) is nontrivial. These defect operators lead to nontrivial symmetry fractionalization and Hall conductance.
\end{itemize}

\subsubsection{Symmetry fractionalization and circuit decomposition}
As implied by the above example, the symmetry domain walls can be nontrivial even when the symmetry operators in the bulk is completely trivial. This happens because there are multiple ways to decompose a given symmetry operator $U$ on a closed surface into a local finite depth circuit. 
Given a symmetry supported on closed submanifold $U(\Sigma_\text{closed})$, if it is a constant depth local unitary circuit, it can be truncated to define on open submanifolds with boundaries $U(\Sigma')$. Such operator creates symmetry defects on the boundary $\partial\Sigma'$.
Importantly, there can be different choices of such truncation with different symmetry defects, and they correspond to {\it different symmetries}, even when the symmetry operators on closed submanifolds are identical.

This happens especially when the symmetries do not act faithfully on a Hilbert space, while the junction of symmetry defects can act by a nontrivial line operator. In other words, this can happen when the 0-form symmetry acts by symmetry fractionalization~\cite{Barkeshli:2014cna,Teo_2015,Tarantino_2016,Benini:2018reh}.
We defined such $U(1)_E\times U(1)_M$ symmetries by the $U^e, U^m$ operators in $\Z_N$ toric code on a lattice, which are the symmetry operators that do not act faithfully on a Hilbert space. One can then define defects of symmetry operators by truncation of a finite depth circuit that generates the symmetry. 

The composition of symmetry operators on submanifold with boundaries for symmetry transformations $g,h,gh\in G$ for symmetry group $G$ produces a string operator on the boundary,
\begin{equation}\label{eqn:compositionstring}
    U_g(\Sigma')U_h(\Sigma')U_{gh}(\Sigma')^{-1}~,
\end{equation}
which corresponds to the action of the junction of symmetry defects on the Hilbert space.
Different truncations of symmetry operators on submanifolds with boundaries give rise to different string operators on the boundaries in the above composition, and they correspond to different choice of symmetry fractionalization class.

In the previous context of symmetry fractionalization, the above truncations are often fixed to a preferred choice, as we describe below:
\begin{itemize}
     \item First of all, once a symmetry operator on a closed surface is given, it fixes the automorphism of local operators in the Hilbert space. One can then use it to define the symmetry defects in the lattice Hamiltonian without reference to circuit decompositions. This fixes a canonical choice of the symmetry defects in lattice models, and choice of a symmetry fractionalization class~\cite{Barkeshli:2014cna}.
    
    \item
    However, the above choice does not always lead to a preferred symmetry fractionalization class; for instance, onsite $U(1)$ symmetry is compatible with trivial gapped phases or FQH states. In these systems, one cannot completely determine the fractionalization class solely from the automorphism of local operators in the bulk. 
    Instead, we fix a truncation of symmetry operators such that $U(\Sigma')$ does not excite the ground state, which can be achieved via quasi-adiabatic insertion of symmetry defects~\cite{Kapustin_2020electric}. Meanwhile, in general 0-form symmetry of (2+1)D gapped phases, we typically do not require the symmetry defects $U(\Sigma')$ to leave the Hamiltonian or ground state energy invariant.\footnote{We remark that such a choice of truncations used in FQH states are not always available. For instance, the symmetry that permutes $e$ and $m$ particles in (2+1)D $\Z_2$ toric code~\cite{Kitaev_2012,Barkeshli:2014cna} with expression in finite depth circuit~\cite{Barkeshli2023codim2, kobayashi2024crosscap} does not have such a truncation: it necessarily creates excitations, and the defect Hamiltonian has modified terms near the location of defects~\cite{shirley:2025unpub}. Even if it does not permute anyon labels, it is unclear if such a preferable truncation exists for the soft symmetry discussed in Ref.~\cite{kobayashi2025softsymmetriestopologicalorders} that acts on junctions of anyons. In generic dimensions, a large class of gauged SPT defects or magnetic defects of finite gauge theory do not admit such truncation, and they have higher-group structure in lattice models that generalizes symmetry fractionalization \cite{Barkeshli:2022edm, Barkeshli:2023bta}.} 
    
\end{itemize}

In summary, there is no generic canonical way to fix a preferred symmetry fractionalization class in previous literature. In this paper, we do not attempt to answer what the generic preferred way to fix a symmetry fractionalization class is. Instead, we allow for non-canonical choices of symmetry defects other than described above, and point out that changing the truncation for symmetry defects can define different symmetries and change the fractionalization classes. 

In particular, for a given onsite symmetry on closed submanifolds on the lattice, it has a canonical choice of onsite decomposition to define defects. With this canonical choice, the Hall conductivity is zero on a local commuting projector model. However, we demonstrated that alternative non-onsite local decomposition of the symmetry gives rise to distinct symmetry fractionalization classes. In such cases, the Hall conductivity can be nonzero.

\section{Higher dimensions: Generalized Hall Conductivities}
\label{sec:generalizedhall3d}

\subsection{Generalized quantum Hall conductivities}

We will generalize the previous discussion of Hall conductivities for ordinary $U(1)$ symmetry in (2+1)D to higher-form symmetry in (3+1)D. The analogue of Hall conductivity between $U(1)$ one-form symmetry and $U(1)$ ordinary symmetry can be described by the transport equation
\begin{equation}
    J^{(1)}=\frac{\xi}{2\pi}\star dB,\quad J^{(2)}=\frac{\xi}{2\pi}\star dA~,
\end{equation}
where $A,B$ are the background one-form and two-form gauge fields for the ordinary and one-form symmetries, respectively, and $J^{(1)},J^{(2)}$ are the corresponding one-form and two-form conserved currents. The transport coefficient $\xi$ will be called the generalized Hall conductivity, and it takes value in real numbers.
The transport can be represented by an effective Chern-Simons like term $\frac{\xi}{2\pi}BdA$.
In particular, the charge of ordinary symmetry in a region is proportional to the flux of the two-form gauge field, with coefficient given by the generalized Hall conductivity $\xi$. 
\begin{equation}\label{eqn:genhall}
    Q^{(0)}=\frac{\xi}{2\pi}\Phi^{(1)}~.
\end{equation}
We remark that similar transports of higher-form symmetry have also been studied in e.g. \cite{Das:2023nwl,thorngren2023higgsU1}.

In a trivially gapped system similar to the integer quantum Hall states, the response action $\frac{\xi}{2\pi}BdA$ should be well-defined by itself, and this requires $\xi$ to be an integer for the action to be invariant under large gauge transformations. When $\xi$ is not an integer, we will say the system has fractional generalized Hall transport.

\subsection{Local commuting projector model}

\begin{figure}[t]
    \centering
    \includegraphics[width=0.7\textwidth]{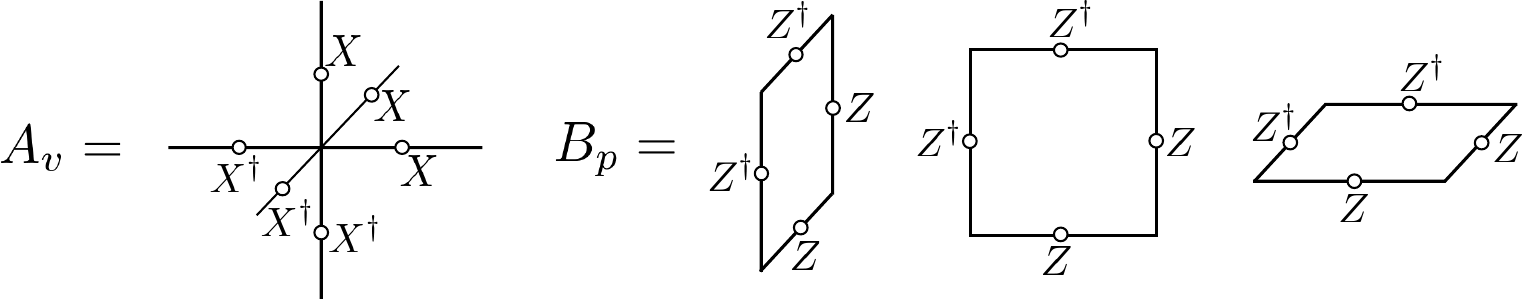}
    \caption{The Hamiltonian of the $\Z_N$ toric code in (3+1)D.}
    \label{fig:3dTC}
\end{figure}

As before, we will consider $\mathbb{Z}_N$ toric code in (3+1)D on cubic lattice, with $\Z_N$ qudit on each edge.
The Hamiltonian is
\begin{equation}
    H=-\sum_v A_v-\sum_p B_p + \text{h.c.}~,
\end{equation}
with the Hamiltonian shown in Figure \ref{fig:3dTC}.
The 1-form $\Z_N$ gauge field is introduced by $e^{2\pi i a(e)/N}:=Z_e$, and the 2-form $\Z_N$ gauge field on the dual plaquettes $\hat e$ is introduced by $e^{2\pi i b(\hat e)/N}:=X_e$. 

The theory has electric and magnetic operators that generate $\mathbb{Z}_N$ two-form symmetry and $\mathbb{Z}_N$ one-form symmetry:
\begin{equation}
    W_e(\gamma)=e^{\frac{2\pi i}{N}\oint_\gamma a},\quad  W_m(\tilde\Sigma)=e^{\frac{2\pi i}{N}\oint_{\tilde\Sigma} b}~,
\end{equation}
where $\gamma$ is a closed loop on the lattice and $\tilde\Sigma$ is a closed surface on the dual lattice.

\paragraph{$U(1)^{(0)}\times U(1)^{(1)}$ symmetry}

We then consider a following combination of symmetries, which are both depth 1 circuits:
\begin{itemize}
\item
$U(1)$ 1-form symmetry generated by
\begin{align}
U^e_\theta(\Sigma) =  \exp\left(\frac{i[\theta]}{N}\int_{\Sigma} d\tilde{a}\right)~,
\end{align}
    \item 
    $U(1)$ 0-form symmetry generated by
    \begin{align}
    V^m_\theta(M) =  \exp\left(\frac{i[\theta]}{N}\int_{M} d\tilde{b}\right)~.
\end{align}
\end{itemize}

\subsection{Protected gapless surface states with symmetries}

\paragraph{Modified Villain $U(1)$ gauge theory in (2+1)D}
The boundary theory with $U(1)^{(0)}\times U(1)^{(1)}$ symmetry can be realized by the Villain $U(1)$ gauge theory \cite{Gross:1990ub,Sulejmanpasic:2019ytl,Gorantla:2021svj} in (2+1)D. This is obtained by starting with $\mathbb{R}$ gauge theory on a square lattice, and then gauging $\Z$ 1-form symmetry. The $\mathbb{R}$ gauge field is described by the Villain $\mathbb{R}$ fields $\{\phi_e,p_e\}$ on the edges of the square lattice, while the $\Z$ 2-form gauge fields are realized by $\{n_f,E_f\}$ on the faces. 

One needs to enforce $\mathbb{R}$ Gauss law on each vertex $v$ (see Figure \ref{fig:doublemodel} for notations),
\begin{align}
    p_{N(v)} + p_{E(v)} - p_{W(v)} - p_{S(v)} = 0~,
    \label{eq:RGausslaw}
\end{align}
as well as the $\Z$ Gauss law on each edge $e$,
\begin{align}
    \exp(-iE_{L(e)}) \exp(-2\pi i p_e) \exp(iE_{R(e)}) = 1~,
    \label{eq:ZGausslaw}
\end{align}
where $L(e), R(e)$ label the faces on the left or right of the edges (with orientations on edges always assigned $\uparrow$ or $\rightarrow$ direction).

Within these Gauss law constraints, there are $U(1)^{(0)}\times U(1)^{(1)}$ symmetries

\begin{align}
    U(1)_{E}^{(1)}(\gamma):\quad  \prod_{e\subset \gamma} \exp(\pm i\theta p_e)~, \qquad  U(1)^{(0)}_{M}:\quad  \prod_{v} \exp(i\theta n_v)~,
\end{align}
where $\gamma$ is a closed loop of the square lattice.

\begin{figure}[t]
    \centering
    \includegraphics[width=\textwidth]{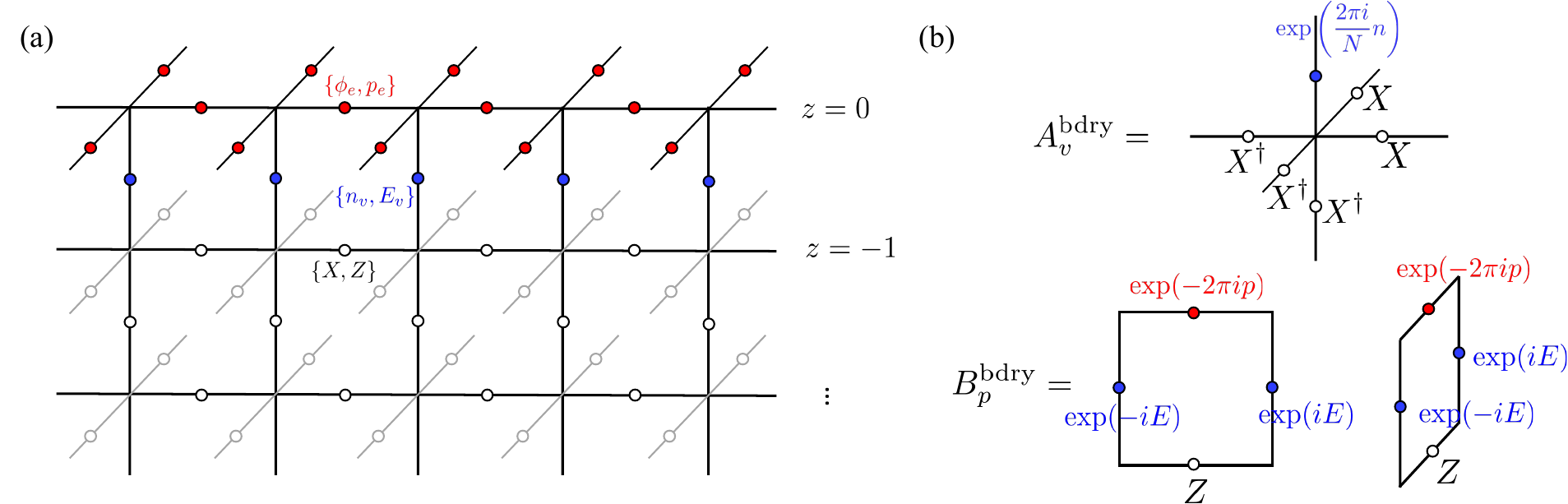}
    \caption{(a): The boundary condition of the (3+1)D $\Z_N$ toric code is given by the Villain $U(1)$ gauge theory. The cubic lattice has a smooth boundary at $z=0$. The $\mathbb{R}$-valued fields $\{\phi_e,p_e\}$ are at the boundary edges $z=0$ (red), and the $\Z$-valued fields $\{n_v,E_v\}$ are at edges $z=-1/2$ (blue). The $\Z$ fields can be labeled by a vertex $v$ at $z=0$. (b): The Hamiltonian that couples the bulk qudits and boundary Villain fields. }
    \label{fig:(2+1)DVillain}
\end{figure}

\paragraph{Symmetric surface state}
The symmetric boundary theory of $\Z_N$ toric code is realized by the Villain $U(1)$ gauge theory in (2+1)D. We again consider the smooth boundary of a 3d cubic lattice, and replace boundary qudits with the Villain fields $\{\phi,p\}$ and $\{n,E\}$ introduced in Sec.~\ref{subsec:1dmV}. See Figure \ref{fig:(2+1)DVillain}.

The Hamiltonian is again given in the form of 
\begin{align}
    H= -\sum_{v\subset \text{bulk}} A_v - \sum_{p\subset\text{bulk}} B_p + 
    H_{\text{mV}}^{\text{bdry}} - g\left(\sum_jA^{\text{bdry}}_{v} +\sum_{j} B^{\text{bdry}}_p\right) + \text{h.c.}~,
\end{align}
where the boundary Villain Hamiltonian has the form of
\begin{align}
    H_{\text{mV}}^{\text{bdry}} = g \sum_{f} (dp)^2|_f + \frac{U_0}{2}\sum_e p_e^2 + \frac{J_0}{2} \sum_v (\phi_{N(v)} +\phi_{E(v)} - \phi_{W(v)} - \phi_{S(v)} + n_v)^2~,
\end{align}
where $dp$ on a boundary face $f=(0123)$ is given by $dp=p_{01} + p_{13} - p_{02} -p_{23}$ (see Figure \ref{fig:doublemodel}). The $g$ term with $g\gg 1$ energetically enforces the $\mathbb{R}$ Gauss law \eqref{eq:RGausslaw} of the $\mathbb{R}$ gauge fields $\phi$ on dual lattice.
The $\Z$ Gauss law \eqref{eq:ZGausslaw} is replaced with the energetic constraint from the boundary Hamiltonian $B_p^{\text{bdry}}$, where the Gauss law is now coupled with the bulk qudit Pauli operator.

The boundary Villain $U(1)$ gauge theory has the following electric $U(1)_E^{(1)}$ and magnetic $U(1)_M^{(0)}$ symmetry:
\begin{align}
    U(1)_{E}^{(1)}: \quad \prod_{e\subset \gamma} \exp(\pm i\theta p_e)~, \qquad  U(1)^{(0)}_{M}:\quad  \prod_{v} \exp(i\theta n_v)~.
\end{align}
The $U(1)_{E}^{(1)}$ symmetry operator is topological due to the 
Gauss law $dp=0$. 

$U_0$, $J_0$ terms in the boundary Villain Hamiltonian are the gauge invariant kinetic terms invariant under $U(1)_{E}^{(1)}\times U(1)^{(0)}_{M}$ symmetries. By enforcing these symmetries, the monopoles are suppressed and the theory becomes gapless.

The electric 1-form symmetry is not $2\pi$ periodic within low energy Hilbert space, but extended by $\Z_N$ in the lattice model;
\begin{align}
    \Z_N^{(1)}\to \tilde U(1)^{(1)}_{E}\to U(1)^{(1)}_{E}~.
\end{align}
where $\Z_N^{(1)}$ is generated by the $\Z_N$ Wilson line of the bulk toric code. The magnetic 0-form symmetry is not extended.

Let us consider the $U(1)^{(0)}\times U(1)^{(1)}$ symmetry operators in the bulk, that end on the boundary.
Such operators solely act on the boundary;
\begin{align}
    U^e_\theta(\Sigma) = \prod_{e\subset \partial \Sigma} \exp(-i\theta p_e)~, \quad V^m_\theta = \prod_{v\subset\text{bdry}} \exp\left(\frac{i\theta}{N}n_v\right)~.
\end{align}
Since $V^m_\theta$ with $\theta=2\pi$ is not identity, $U(1)^{(0)}$ symmetry is no longer $2\pi$ periodic, but becomes $2\pi N$ periodic. 

The action of $U^e_\theta$ is identified as the electric 1-form symmetry of the Villain $U(1)$ gauge theory, which is extended by $\Z_N$ in the lattice model;
\begin{align}
    \Z_N^{(1)}\to \tilde U(1)^{(1)}\to U(1)^{(1)}~.
\end{align}

\paragraph{Anomaly of boundary theory} In the bulk, the $U(1)^{(0)}$ ordinary and $U(1)^{(1)}$ one-form symmetries act on the theory through symmetry fractionalization with the $\mathbb{Z}_N$ one-form and $\mathbb{Z}_N$ two-form symmetries.
The $U(1)^{(0)}\times U(1)^{(1)}$ symmetry has a mixed topological response: the low energy theory describes $\mathbb{Z}_N$ gauge theory that couples to the background two-form gauge fields $A_E$ for the $U(1)^{(1)}$ one-form symmetry and background one-form gauge field $A_M$ for the $U(1)$ ordinary symmetry,
\begin{equation}
    \frac{N}{2\pi}adb+\frac{1}{2\pi}adA_E+\frac{1}{2\pi}bdA_M~,
\end{equation}
where $a,b$ are the one-form gauge field and two-form dual gauge fields, respectively. Integrating out $a,b$ gives trivial fractional generalized Hall response: denote the $U(1)$ two-form current by $J_E=\star da/(2\pi)$ and $U(1)$ one-form current by $J_M=\star db/(2\pi)$, they are related to the background fields as
\begin{equation}
    J_E=-\frac{(1/N)}{2\pi}\star dA_M,\quad J_M=-\frac{(1/N)}{2\pi}\star dA_E~.
\end{equation}
The generalized Hall conductivity is thus
\begin{equation}
    \xi=-\frac{1}{N}\text{ mod }1~.
\end{equation}
The transport can be represented by the effective action $-\frac{(1/N)}{2\pi}A_EdA_M$.

On the boundary, the $U(1)^{(0)}$ symmetry is extended to be $N$-fold covering $\tilde U(1)^{(0)}$, thus we replace $A_M=N\tilde A_M$. Thus the boundary has mixed anomaly for $\tilde U(1)^{(0)}\times U(1)^{(1)}$ symmetry, given by properly quantized mixed Chern-Simons term $N\times (-1/N)=-1$:
\begin{equation}
    -\frac{1}{2\pi}A_E d\tilde A_M~.
\end{equation}
This is consistent with the mixed anomaly of $U(1)^{(1)}\times U(1)^{(0)}$ in the boundary (2+1)D $U(1)$ gauge theory.

\subsection{Non-vanishing generalized Hall conductance}

\subsubsection{Surface currents}

The generalized Hall conductance can be computed from the commutator of charge density operators on the boundary. In our case, the charge densities of the two $U(1)$ symmetries are $p_e$ and $\frac{1}{2\pi}\left(d\phi(v)+n_v\right)$, respectively, where $d\phi(v)=\phi_{N(v)}+\phi_{E(v)}-\phi_{W(v)}-\phi_{S(v)}$ is the flux of $d\phi$ evaluated on the plaquette on the dual lattice that is dual to vertex $v$.
The commutator gives
\begin{equation}
    [\frac{1}{2\pi}\left( d\phi(v)+n_v\right),p_e]=\frac{i}{2\pi}\left(\delta_{e,N(v)}+\delta_{e,E(v)}-\delta_{e,W(v)}-\delta_{e,S(v)}\right)~.
\end{equation}
Thus the two $U(1)$ symmetries have a mixed anomaly with unit coefficient. In terms of the bulk $U(1)$ symmetries related by $N$-fold covering, this indicates the mixed Hall conductivity $1/N$ between $U(1)_E\times U(1)_M$.

\subsubsection{Generalized many-body Chern number}

An analogue of the many-body Chern number can again be used to characterize the generalized Hall conductivity. Let us consider a system on a cubic lattice with periodic boundary conditions on a 3d torus $S^1_x\times S^1_y\times S^1_z$. We introduce twisted boundary conditions for the $U(1)^{(0)}\times U(1)^{(1)}$ symmetry, such that $U(1)^{(0)}$ symmetry has a holonomy $\theta_x$ along $x$ direction, and $U(1)^{(1)}$ symmetry has a holonomy $\theta_{yz}$ along the $yz$ plane. Let us denote the twisted Hilbert space by $H[\theta_x, \theta_{yz}]$. In our example, this again defines a family of gapped commuting projector models parametrized by $T^2$; $(\theta_x,\theta_{yz})$. The many-body Chern number is then defined in the exactly same fashion as the standard case.
Following the logic of Sec.~\ref{subsec:chernnumber}, it is again computed by the commutation relation of the $2\pi$ flux insertion operators on the ground state $\ket{\Psi[0,0]}$ without twist,
\begin{align}
    \exp(2\pi i C) = \bra{\Psi[0,0]}V^\dagger_{2\pi,x}V^\dagger_{2\pi,yz}V_{2\pi,x}V_{2\pi,yz}\ket{\Psi[0,0]} =e^{2\pi i/N}~,
\end{align}
where $V_{2\pi,yz} = \prod_{\text{line}}Z^\dagger$ is a line operator extended in $x$ direction, inserting $2\pi$ holonomy of $U(1)^{(1)}$ symmetry along $yz$ plane. $V_{2\pi,x}= \prod_{\text{surface}} X^\dagger$ is a surface operator extended in $yz$ direction, inserting $2\pi$ holonomy of $U(1)^{(0)}$ symmetry along $x$ direction.
This shows that the many-body Chern number becomes $1/N$ mod 1.

\subsubsection{Generalized flux insertion}

The generalized Hall conductance can be computed locally using (\ref{eqn:genhall}) on sphere space,
\begin{equation}
 Q^{(0)}=\frac{\xi}{2\pi}\Phi^{(1)}~.   
\end{equation}
Consider junction of three symmetry defects for the one-form symmetry, with transformation $\theta_1,\theta_2,\theta_3=[\theta_1+\theta_2]$. At the center of the junction  there is magnetic flux $\Phi^{(1)}$ of the two-form gauge field for the one-form symmetry:
\begin{equation}
    \Phi^{(1)}=\theta_1+\theta_2-[\theta_1+\theta_2]~.
\end{equation}
The generalized Hall conductivity relates the flux and the electric charge $Q^{(0)}=\int d\tilde b/N$ of the ordinary symmetry.

Since we only care about the fractional part of the charge, we can compute the value of $e^{2\pi iQ^{(0)}}$ for region enclosing the center of the junction.
In order to compute the charge at the center of the junction, we fold the symmetry defects similar to Figure \ref{fig:junction}, then the only excitation is at the center of the folded junction, created by the string operator 
\begin{equation}
    W_e^{\left(\theta_1+\theta_2-[\theta_1+\theta_2]\right)/2\pi}=W_e^{\Phi^{(1)}/2\pi}~.
\end{equation}
This gives the braiding phase $e^{{2\pi i\over N}\frac{\theta_1+\theta_2-[\theta_1+\theta_2]}{2\pi}}=e^{\frac{i}{N}\left(\theta_1+\theta_2-[\theta_1+\theta_2]\right)}$, and thus the charge is
\begin{equation}
    Q^{(0)}=\frac{1}{2\pi N}\left(\theta_1+\theta_2-[\theta_1+\theta_2]\right)=\frac{1}{N}\frac{\Phi^{(1)}}{2\pi}\text{ mod }1~.
\end{equation}
Thus we conclude that the system has non-vanishing fractional generalized Hall conductivity:
\begin{equation}
    \xi=\frac{1}{N}\text{ mod }1~.
\end{equation}

\subsection{Comments on Fractionalization}

These $U(1)$ ordinary and one-form symmetries have nontrivial symmetry fractionalization. To see this, we can compare the transformations $U_\theta(R)$ for $\theta,\theta'$ and $[\theta+\theta']$ on a two-dimensional region $R$ with boundary:
\begin{equation}
    U_\theta(R)U_{\theta'}(R)=U_{[\theta+\theta']}(R) W_e(\partial R)^{\eta(\theta,\theta')}.
\end{equation}
Thus the magnetic flux loop excitations created by open $W_m^{q_m}$, which braids with $W_e$, carry fractional $U(1)$ one-form charge $\frac{q_m}{N}$. Such fractionalized $U(1)$ one-form symmetry is also discussed in \cite{Hsin:2019fhf}.

Similarly, we can compare the transformation $V_\theta(R')$ for $\theta,\theta'$ and $[\theta+\theta']$ on a three-dimensional region $R'$ with boundary:
\begin{equation}
    V_\theta(R')V_{\theta'}(R')=V_{[\theta+\theta']}(R) W_m(\partial R')^{\eta(\theta,\theta')}.
\end{equation}
Thus the electric charge excitation created by open $W_e^{q_e}$, which braids with $W_m$, carry fractional ordinary $U(1)$ charge $\frac{q_e}{N}$.

\section{Gapless Edge Modes from Generalized Hall Conductivity}
\label{sec:protectededge}

In this section we will provide field theory arguments for symmetry-enforced gapless boundary for the generalized Hall response of generalized symmetries.

Consider the topological response of $U(1)$ $q_i$-form symmetries described by
\begin{equation}
S_\text{response}=    \sigma\int A_{i} \prod_{j\neq i}\frac{dA_j}{2\pi}  ~,
\end{equation}
where $A_i$ is the background for $U(1)$ $(q_i-1)$-form symmetry and it is a $q_i$-f-form gauge field, and $\sigma$ is the topological transport.
The response is such that the total spacetime dimension $D$ satisfies
\begin{equation}
    D=q_i+\sum_{j\neq i} (q_{j}+1)~.
\end{equation}
When the system is an SPT, $\sigma$ is quantized to be an integer.  

Let us argue that the boundary without explicitly breaking the $U(1)$ symmetries must be gapless. For this purpose, we can focus on the case that $\sigma$ is an integer by taking covering of the $U(1)$ symmetries.
We will provide two arguments: the first argument shows the boundary has non-constant long-range correlation function, and the second argument shows the boundary has vanishing partition function.

We remark that the discussion also applies to continuous $O(2)/\mathbb{Z}_2^{\cal C}$ non-invertible cosine symmetry where a $\mathbb{Z}_2$ complex conjugation action for some of the $U(1)$ symmetry is gauged: as long as the transport coefficient $\sigma$ is invariant, the discussion here carries through.

\subsection{Gapless boundary from long-range correlation}

Consider $U(1)$ transformation of $A_i\rightarrow A_i+d\alpha$, the transport implies that the boundary acquires an anomalous transform
\begin{equation}
    e^{i\sigma\int \alpha\prod_{j\neq i} {dA_{j}\over 2\pi} }~.
\end{equation}
This implies that the currents for these $U(1)$ symmetries have the correlation function in momentum space\footnote{
In path integral formalism, the background gauge field couples to current as $e^{i\int A_\mu j^\mu}$. Under gauge transformation $A_\mu \rightarrow A_\mu+\partial_\mu\alpha$, this gives $e^{i\int A_\mu j^\mu-i\int \alpha \partial_\mu j^\mu}$, and the partition function transforms by the anomalous phase $e^{i\int \alpha \omega(A)}$ for $\omega(A)=\sigma \prod dA_j/2\pi$.
Taking functional derivative with respect to $\alpha$ and then set $\alpha=0$ gives $\langle \partial_\mu j^\mu \rangle=-\omega(A)$.
}
\begin{equation}
    k_{\mu_1}\langle  J^{\mu_1\cdots \mu_{q_i}}_i(k)\prod_j J_j^{\nu_1\cdots \nu_{q_j}}(p_j)\rangle = {\sigma\over 2\pi} \epsilon^{\mu_2\cdots,\mu_{q_i},\{\lambda_j,\nu_1,\cdots,\nu_j\}}\prod p_{\lambda_j}~,
\end{equation}
where $k=-\sum_j p_j$. From this contact term correlation function one can solve for the correlation function without the derivative $k_{\mu_1}$:
\begin{equation}
\langle  J^{\mu_1\cdots \mu_{q_i}}_i(k)\prod_j J_j^{\nu_1\cdots \nu_{q_j}}(p_j)\rangle = {\sigma\over 2\pi} \epsilon^{\mu_2\cdots,\mu_{q_i},\{\lambda_j,\nu_1,\cdots,\nu_j\}}\frac{k^{\mu_1}\prod p_{\lambda_j} }{k^2}~,
\end{equation}
where we can average the right hand side with respect to exchanging $\mu_1$ with $\mu_{j}$ for $j=2,\cdots,q_i$.
The right hand side is not analytic in every momentum $p_i$, and thus the correlation function for the currents without derivative is not a contact term and non-vanishing at separate points. This indicates that the boundary is gapless as long as $U(1)$ symmetry is not explicitly broken. Note that the argument works for any nonzero $\sigma$.

\paragraph{Example: ordinary quantum Hall transport}

It is instructive to consider the example of quantum Hall transport for $U(1)\times U(1)$ 0-form symmetry.
The boundary $U(1)$ currents satisfies
\begin{equation}
    \langle J_1(z)J_2(w)\rangle=\frac{\sigma/2\pi}{(z-w)^2}=\frac{\sigma}{2\pi} \partial_z \frac{1}{z-w}~.
\end{equation}
If we apply $\partial_{\bar z}$, this gives
\begin{equation}
    \langle \partial_{\bar z}J_1(z)J_2(w)\rangle=\frac{\sigma/2\pi}{(z-w)^2}=\sigma\partial_z \delta^2(z-w)~.
\end{equation}

\subsection{Gapless boundary from vanishing partition function}

For simplicity, let us focus on ``quadratic'' topological response
\begin{equation}
    \frac{\sigma}{2\pi}A_{i}dA_{j}~,
\end{equation}
where $D=q_i+q_j+1$. Let us take the boundary to be $S^{q_{i}-1}\times S^{q_j+1}$.
Again we perform the transformation $A_i\rightarrow A_i+d\alpha$, with $\alpha|=\alpha_0 \text{vol}_{S^{q_i-1}}$ for constant $\alpha_0$. This is a global $(q_i-1)$-form transformation on the boundary.
We also turn on backgrounds $dA_j=2\pi m_j\text{Vol}_{S^{q_j+1}}$ for integer $m_j$.
Since $\alpha$ is a global symmetry transformation that leaves $A_i$ invariant on the boundary, the boundary partition function satisfies
\begin{equation}
    Z[A_i,A_j]=Z[A_i,A_j]e^{i\alpha_0 m_j}~.
\end{equation}
For $\alpha_0\not\in {2\pi\over \prod m_j}\mathbb{Z}$, this implies that the partition function vanishes $Z=0$. We will take $A_i=0$ in the following, with nontrivial background $A_j$ only on $S^{q_j+1}$.

Let us show that this is not possible if the boundary is gapped with finite number of vacua.
We will argue by contradiction. Suppose the boundary is gapped with finite number of vacua. 

The boundary partition function is partition function on  $S^{q_i-1}\times S^{q_j+1}$.
Let us compactify the gapped theory on $S^{q_i-1}$ that does carry any flux of the background fields.
This gives a compactified TQFT also with finite number of vacua, with the same partition function $Z=Z_{S^{q_j+1}}$. 

If there is no flux of the $q_j$-form gauge field $A_j$, the partition function on $S^{q_j+1}$ of the compatified TQFT would be the sum of the norm of the states on hemisphere and thus nonzero:
\begin{equation}
    \text{No flux }m_j=0:\quad Z_{S^{q_j+1}}=\sum_i \langle i|i\rangle\neq 0~,
\end{equation}
where $|i\rangle$ are the states in the compactified TQFT on hemisphere $D^{q_j+1}$.

In the presence of flux $m_j$ on the $(q_j+1)$-dimensional sphere, i.e. $\int_{S^{q_j+1}}dA_j=2\pi m_j$, the partition function becomes the sum of matrix elements:
\begin{equation}
    Z_{S^{q_j+1}}=\sum_i \langle i|  T_j^{m_j}|i\rangle~,
\end{equation}
where the operator $T_j$ glues the two hemispheres together by the transformation $A_j\rightarrow A_j+ d\tau$, where the $(q_j-1)$-form higher Berry connection $\tau$ satisfies $d\tau=\text{Vol}_{S^{q_j}}$, along the equator $S^{q_j}$ of $S^{q_j+1}$. This is the clutching construction for the bundle of $A_j$ with magnetic flux.

The operators $T_j^\ell$ for different integer power $\ell$ generates $\mathbb{Z}$ group. Since the gapped system and its compactification on finite dimensional spheres without fluxes has finite number of vacua, the operator acting on the compactified TQFT Hilbert space as a finite dimensional representation of the infinite $\mathbb{Z}$ group, and thus the representation is not faithful. This implies that there exists an integer $N_j$ such that $T^{N_j}=1$ on the Hilbert space.
Now, we take the background to have flux $m_j\in N_j\mathbb{Z}$, then $T_j^{m_j}=1$ acting on the compactified TQFT Hilbert space. Thus the partition on the $S^{q_j+1}$ with the background $A_j$ is nonzero:
\begin{equation}
    \text{For flux }m_j\in N_j\mathbb{Z}:\quad Z_{S^{q_j+1}}=\sum_i \langle i|  T_j^{m_j}|i\rangle=\sum_i \langle i|i\rangle\neq 0~.
\end{equation}

On the other hand, we can choose $\alpha_0\not\in \frac{2\pi}{m_j}\mathbb{Z}$ to show that the partition function vanishes in the presence of such flux. Thus we conclude that the boundary must be gapless. 
The argument can be generalized to case of topological response involves more than two $U(1)$ higher-form gauge fields.

\paragraph{Gapped system with vanishing partition function}

To see that gapped systems can have vanishing partition function, consider $\mathbb{Z}_2$ gauge theory in (1+1)D enriched by $U(1)$ symmetry, with action
\begin{equation}
    \frac{2}{2\pi}ad\phi+\frac{1}{2\pi}\phi dA~,
\end{equation}
where $a,\phi$ are dynamical fields.
The coupling to background gauge field $A$ tells us that if the magnetic flux is not an even integer multiple of $2\pi$, the partition function on $S^2$ vanishes. On the other hand, if the magnetic flux is an even integer of $2\pi$, then the flux decouples from the $\mathbb{Z}_2$ gauge theory, and the partition function on $S^2$ is nonzero.

\section{Outlook}
\label{sec:outlook}

There are several future directions. First, while in the continuum limit the $U(1)$ ordinary and higher-form symmetries are non-anomalous in the local commuting projector models, it is not clear how to gauge the symmetries since there is no local commuting charge operator that can be used to impose local Gauss law. Since gauging these symmetries would give rise to Chern-Simons like theories on the lattice (see e.g. \cite{DeMarco:2019pqv,Jacobson:2024hov} for previous constructions), it can be interesting to carry out the gauging procedure on the lattice.

Second, we can ask the consequences of generalized Hall conductivities to the circuits that generate the symmetries. This is similar to indices that relate to anomalies of the symmetries and whether the symmetries are onsite \cite{Else:2014vma,Seifnashri:2025vhf,Kapustin:2025nju}.
The $U(1)$ symmetry gives a homomorphism to Quantum Cellular Automata (QCA), $\rho:U(1)\rightarrow \text{QCA}$. In particular, we do not require $U(1)$ symmetry has local commuting charge operators.
In 2d space the QCA is a circuit, so we can truncate the homomorphism on half space to give $\rho_L,\rho_R$ separated by interface $\gamma$. Let us fix a choice of $\rho_L$.~\footnote{For onsite symmetries, $\rho_L$ is usually chosen using a sum of ``dressed'' version of local charge operators, which are local charges that do not excite the ground state \cite{Kapustin_2020electric}. In the current discussion, this would enable nontrivial symmetry fractionalization (i.e., $\alpha$) with onsite symmetries.} Then one can compute $\rho_L(\theta_1)\rho_L(\theta_2)\rho_L(\theta_1+\theta_2)^{-1}$, which is a 1d QCA supported on the interface $\gamma$:
\begin{equation}
\alpha_\gamma(\theta_1,\theta_2):=\rho_L(\theta_1)\rho_L(\theta_2)\rho_L(\theta_1+\theta_2)^{-1}\in \text{QCA}(\gamma)~.
\end{equation}
Suppose that the GNVW index \cite{Gross_2012} of the 1d QCA $\alpha_\gamma(\theta_1,\theta_2)$ is trivial, then $\alpha_\gamma(\theta_1,\theta_2)$ is also a circuit and one can define truncated operators.
Then one can compute the statistics using the truncated operators using the processes in \cite{Kawagoe_2020,Kobayashi:2024dqj}. This extracts the fractional part of the Hall conductance $\sigma$ mod 1, even if the $U(1)$ symmetry does not have local commuting charge operators. Note that the procedure yields trivial fractional part for the Hall conductance of Chern insulators with onsite $U(1)$ symmetry. We will leave detailed discussion to a separate work.

Third, systems with fractional generalized Hall transports are expected to exhibit long-range entanglement, as short-range entanglement systems will have integer quantized transport coefficients. It would be interesting to quantify this on the lattice using methods such as in \cite{Kapustin_2020,Hsin:2025pqb}.

Finally, while we focus on $U(1)$ ordinary or higher-form symmetries, it would be interesting to generalize the discussion to other continuous symmetries including the transports of continuous non-invertible symmetries, as discussed in e.g. \cite{Hsin:2024aqb,Hsin:2025ria}.

\section*{Acknowledgment}
We thank Maissam Barkeshli, Meng Cheng, Dominic Else, Nathan Seiberg, Sahand Seifnashri, Nikita Sopenko and Carolyn Zhang for discussions. We thank Maissam Barkeshli for comments on a draft.
 R.K. is supported by the U.S. Department of Energy, Office of Science, Office of High Energy Physics under Award Number DE-SC0009988 and by the Sivian Fund at the Institute for Advanced Study. P-S.H. is supported by Department of Mathematics, King's College London. R.K. and P.-S.H. thank the Kavli Institute for Theoretical Physics for hosting the program “Generalized Symmetries in Quantum Field Theory: High Energy Physics, Condensed Matter, and Quantum Gravity” in 2025, during which part of this work was completed. This research was supported in part by grant no. NSF PHY-2309135 to the Kavli Institute for Theoretical Physics (KITP).

 \appendix

\section{Review of Modified Villain Formalism}
\label{app:modifiedVillain}

In this appendix we will review the modified Villain formalism in \cite{Gross:1990ub,Sulejmanpasic:2019ytl,Gorantla:2021svj} for $q$-form $U(1)$ gauge theories in $D$ spacetime dimensions using the Hamiltonian language.

Consider discretized time and space, with action
\begin{equation}
    S=\frac{\beta}{2}\sum \left(da^{(q)}-2\pi n^{(q+1)}\right)^2+i\sum \tilde a^{D-q-2} dn^{(q+1)}~,
\end{equation}
where the sum is over suitable cells on spacetime lattice.
The gauge transformations are
\begin{align}
 &   a^{(q)}\rightarrow a^{(q)}+d\lambda^{(q-1)}+2\pi k^{(q)},\quad  n^{(q+1)}\rightarrow n^{(q+1)}+dk^{(q)}\cr 
 &\tilde a^{(D-q-2)}\rightarrow \tilde a^{(D-q-2)}+d\tilde \lambda^{(D-q-3)}+2\pi \tilde k^{(D-q-2)}~,
\end{align}
where $a^{(q)},\tilde a^{(D-q-2)}$ as well as the transformation parameters $\lambda^{(q-1)},\tilde \lambda^{D-q-3}$ take value in $\mathbb{R}$, while $n^{(q+1)}$ and the transformation parameters $k^{(q)},\tilde k^{D-q-2}$ take value in $\mathbb{Z}$.
The purpose of $\tilde a^{(D-q-2)}$ is to suppress the flux of $n^{(q+1)}$.

Consider discretized space with continuous time. The Hamiltonian can be obtained from the action canonically as follows. The conjugate momentum of $a^{(q)}$ is $p=\beta(\partial_0 a^{(q)}-2\pi n^{(q+1)})$. Since the $\tilde a$ term has first order derivative in the action, they do not appear in the Hamiltonian, but instead as constraint after integrating out the non-dynamical variables. Thus,
\begin{equation}
    H=\frac{1}{2\beta}\sum p^2+ \frac{\beta}{2}\sum \left(\vec{\nabla} a^{(q)}-2\pi n^{(q+1)}\right)^2~,
\end{equation}
where the sum is over suitable cells on space lattice.
Note that there is no electric field term for the $\mathbb{Z}$ gauge field. Example of such Hamiltonian is discussed in \cite{Cheng:2022sgb}.

\bibliographystyle{utphys}
\bibliography{biblio.bib}
\end{document}